\documentstyle[12pt,aaspp4,psfig]{article}
\font\cap=cmcsc10
\tightenlines

\newcommand\hi{\noindent \hangindent=2.5em}
\newcommand\et{{\it et\thinspace al.}}    

\newcommand\kpc{{\rm\,kpc}}

\newcommand\kmsec{{\rm\,km/s}}
\newcommand\kms{\kmsec}

\newcommand\surfb{{\rm\,mag/arcsec^2}}

\newcommand\aanda{{\rm Astr.~Ap.}, }     
\newcommand\aasup{{\rm Astr.~Ap.~Suppl.}, }     

\begin{document}

\title{A Structural and Dynamical Study of Late-Type, Edge-On Galaxies:
I. Sample Selection and Imaging Data}

\author{Julianne J. Dalcanton\altaffilmark{1}}
\affil{Department of Astronomy, University of Washington, Box 351850,
Seattle WA, 98195 \\ \& \\
Observatories of the Carnegie Institution
        of Washington, 813 Santa Barbara Street, Pasadena CA, 91101}

\author{Rebecca Bernstein\altaffilmark{2,3}}
\affil{Observatories of the Carnegie Institution
        of Washington, 813 Santa Barbara Street, Pasadena CA, 91101}

\altaffiltext{1}{e-mail address: jd@astro.washington.edu}
\altaffiltext{2}{e-mail address: rab@ociw.edu}
\altaffiltext{3}{Hubble Fellow}
  
\begin{abstract}

  We present optical ($B$ \& $R$) and infrared ($K_s$) images and
  photometry for a sample of 49 extremely late-type, edge-on disk
  galaxies selected from the Flat Galaxy Catalog of Karenchentsev et
  al.\ (1993).  Our sample was selected to include galaxies with
  particularly large axial ratios, increading the likelihood that the
  galaxies in the sample are truly edge-on.  We have also concentrated
  the sample on galaxies with low apparent surface brightness, in
  order to increase the representation of intrinisically low surface
  brightness galaxies.  Finally, the sample was chosen to have no
  apprarent bulges or optical warps so that the galaxies represent
  undisturbed, ``pure disk'' systems. The resulting sample forms the
  basis for a much larger spectroscopic study designed to place
  constraints on the physical quantities and processes which shape
  disk galaxies.

  The imaging data presented in this paper has been painstakingly
  reduced and calibrated to allow accurate surface photometry of 
  features as faint as $30\surfb$ in $B$ and $29\surfb$ in $R$ on
  scales larger than 10$\arcsec$.  Due to limitations in sky
  subtraction and flat fielding, the infrared data can reach only to
  $22.5\surfb$ in $K_s$ on comparable scales.  As part of this work,
  we have developed a new method for quantifying the reliability of
  surface photometry, which provides useful diagnostics for
  the presence of scattered light, optical emission from infrared
  cirrus, and other sources of non-uniform sky backgrounds.

\end{abstract}


\section{Introduction}                  \label{intro}

Basic physics must be responsible for final differentiation of
galactic structure.  Mass, angular momentum, density, environment, and
metallicity all must contribute to the shape and relative proportions
of the disk and spheroid structures observed today.  While the
tremendous diversity of the galaxy population suggests a bewildering
level of complexity in the {\it details} of galaxy formation, the
existence of broad patterns, such as the Fundamental Plane or the
Tully-Fisher relation, gives some hope that the overall structure of
galaxies are controlled by large scale physics, and thus can be
explained and constrained with observation.

Disk galaxies represent some of the best possible laboratories for
exploring the physics which controls galaxy formation.  Spiral disks
are less corrupted by dissipation and angular momentum transport than
comparable elliptical galaxies, and thus they better preserve the
initial conditions from which they were formed.  Likewise, the disks
of spiral galaxies extend far out into their dark matter halos, and
thus can be used to probe the shape and extent of the
accompanying dark matter, which in turn places strong constraints on
theories of dark matter and structure formation.  Finally, only in
spirals can we directly observe galaxy formation {\it in process},
particularly among late type spiral disks and low surface brightness
galaxies, which, from their colors, IR surface brightnesses, and gas
content, seem to be forming stars for nearly the first time.

To place observational constraints upon the process of galaxy
formation, we have begun a comprehensive program to study the
dynamics, gas content, metallicity, and stellar populations of a
sample of late-type, bulgeless disk galaxies.  This population
forms a structurally uniform sample, allowing us to isolate changes in
the physical properties of the galaxies (i.e. mass, angular momentum,
etc.) independent of changes in morphology.  By avoiding systems with
bulges, we also limit the degree to which the baryonic component of
the galaxy may have been affected by dissipation or angular momentum
transport during formation.

We have selected these galaxies from a large catalog of edge-on
galaxies, described below in \S\ref{samplesec}.  By selecting the
galaxies edge-on, we can ensure that the galaxies are free of strong
warps, which could indicate a recent interaction or a non-equilibrium
configuration.  The galaxies in our sample should therefore be well
relaxed and largely undisturbed.  Furthermore, while the edge-on view
of a galaxy can clearly identify it as a disk, it disguises the
face-on morphology, masking the presence or absence of spiral arms,
bars, or star formation regions.  Thus, the selection of an edge-on
sample should be unbiased with respect to these transient features.
Finally, the edge-on orientation of these galaxies allows direct study
of their vertical structure.  The vertical structure of galactic disks
contains information on the balance between the surface density of the
disk, the vertical velocity dispersion of the stars, and the density
structure of the halo.  Thus this sample will provide constraints
(albeit highly interdependent constraints) on the internal dynamics of
the disk and the flattening of the halo (Spitzer 1942, van der Kruit
\& Searle 1982, Bahcall 1984, Zasov et al.\ 1991, Dove \& Thronson
1993, Olling 1995, van der Kruit \& de Grijs 1999).

Most importantly, by selecting galaxies edge-on, we allow ourselves to
sample a wider range of disk surface brightnesses.  Low surface
brightness galaxies tend to have low internal extinction (de Blok \&
van der Hulst 1998, McGaugh 1994), and thus there is a substantial
enhancement in their apparent surface brightness when viewed edge-on.
Selecting edge-on galaxies therefore admits galaxies which would
otherwise be too low surface brightness to detect or to study
(particularly in the infrared).  As the dynamics of low surface
brightness galaxies tend to be dominated by the dark matter halo
(Swaters et al.\ 2000, de Blok \& McGaugh 1997; although see van den
Bosch et al.\ 2000), these systems make ideal probes of the halo's
density structure.  

Observationally, we are studying these galaxies through a combination
of optical and infrared imaging (to constrain the mass of the stellar
disk and the structure of the galaxy, and to roughly limit the current
stellar populations and dust content), high-resolution long-slit
spectroscopy of the H$\alpha$ line and nearby [NII] and [SII] doublets
(to study the internal dynamics of the disk and limit the abundance
of the gas phase), low-resolution spectroscopy of the region between
H$\beta$ and H$\alpha$ (to map internal extinction across the disk),
and HI observations (to constrain the total mass and distribution of
atomic gas in the disk).  In this first paper, we present the sample,
the imaging data, and the resulting photometry, with a complete
analysis of the errors and uncertainties.  We delay a full analysis of
the galaxy colors and structural properties until a later paper in the
series.


\section{Sample Selection}                              \label{samplesec}

The sample was selected from the Flat Galaxy Catalog (FGC) of
Karanchentsev et al.\ (1993), a catalog of 4455 edge-on galaxies with
axial ratios greater than 7, and major axis lengths of $> 0.6\arcmin$.
The FGC was originally selected by visual inspection of the O POSS
plates in the north ($\delta>-27\deg$) and the J films of the ESO/SERC
survey in the south ($\delta < -17\deg$). Galaxies from the ESO plates
are known as the FGCE, and have slightly different properties due to
small differences in the plate material.  From the combined FGC/FGCE
catalog, we selected galaxies which appeared both bulgeless and low
surface brightness on the Digitized Sky Survey (DSS), and which showed
no signs of inclination (major-to-minor axis ratio $a/b>8$) or
interaction.  Unfortunately, our selection criteria were not uniformly
successful, as the images in \S\ref{magsec} and Figure
\ref{contourfig} will show.  Due to the low resolution of the DSS,
$\lesssim$10\% of the galaxies which met the original selection criteria
showed small bulges which were not apparent on the DSS, or dust
lanes which masked a high-surface brightness disk (e.g.\ FGC 446, FGC
1043, FGC 1440, \& FGC E1371).  As these galaxies will be useful for
some aspects of our extended scientific program, we have retained them
in the sample, but treated them separately when appropriate.  One
galaxy which we had originally chosen for the survey, FGC E1550,
showed a pronounced integral--sign--shaped warp in our initial $R$ band
imaging, and was removed from the sample.  The final sample is listed
in Table \ref{positiontable} along with positions and orientations as
given in the FGC.

The resulting distribution of morphological types and surface
brightness classes (I=high surface brightness, IV=low surface
brightness), both as given by the FGC, are plotted in Figure
\ref{sbtypefig}, along the with the distributions for the entire FGC
catalog.  Clearly our sample is biased towards later types and lower
surface brightnesses than the FGC as a whole.  We have also plotted
the distributions of blue axial ratios for our sample in Figure
\ref{axialratiofig}.  Our subsample has a higher mean axial ratio than
the catalog as a whole, betraying our selection bias for the most nearly
edge-on galaxies.  It may also reflect our choice of late
morphological types; an analysis of the FGC by Kudrya et al.\ (1994)
shows that the galaxies in the FGC become progressively thinner with
later Hubble types, with the limiting axial ratio varying from
$(a/b)_{max}=14.1$ for Sb galaxies to $(a/b)_{max}=27.0$ for Sd's.  This
trend towards intrinsically thinner galaxies with increasing
Hubble type is also seen by de Grijs (1998). The
apparent bias towards large axial ratios may also result from our
selection of lower surface brightness galaxies. LSBs are known to have
larger disk scale lengths than normal galaxies with similar rotation
speeds (Zwaan et al 1995), and consequently may have larger axial
ratios as well.

Many of the galaxies in our sample were previously observed in single
dish HI observations with Arecibo.  For our 8h$<$RA$<$19h sample
(spring observing season), we concentrated on those galaxies which had
HI detections and which were relatively nearby ($V \lesssim 5000\kms$),
giving us better spatial resolution for both imaging and spectroscopy.
During the fall, slightly more than half (18 of 32) of our sample
galaxies had existing HI observations.  Overall, 75\% of the galaxies
in our survey have published single dish HI observations.  When
available, the heliocentric velocity and the corrected line width at
50\% peak flux ($W_{50,c}$) are listed in Table \ref{positiontable}.
The majority of these are from a large survey of FGC galaxies observed
at Arecibo by Giovanelli et al.\ (1997).  These are supplemented with
measurements for FGC 164 from Schneider et al.\ (1990), and for FGC
349 from Haynes et al.\ (1997).

In addition to HI observations, a very small number of the galaxies in
our sample (FGC 446, FGC 1043, and FGC 2217) were detected as part of
the IRAS Faint Source Catalog (Moshir et al.\ 1990; F03422+0544,
F10131+0734, 18356+1729 respectively).  In order of increasing 60$\mu$
flux, FGC 1043 is detected in both 60$\mu$ and 100$\mu$ bands, with
flux of 0.23~Jy and 0.65~Jy, respectively; FGC 446 is only detected at
60$\mu$, with flux of 0.38~Jy; and FGC 2217 is detected at 25$\mu$ and
60$\mu$, with flux of 0.17~Jy and 1.25~Jy, respectively.  All of the
detections and upper limits are consistent with spectra which rise in
$\nu f_\nu$ towards 100$\mu$.

While our sample galaxies are exceptionally useful probes of the
properties of galaxies over a spectrum of mass and surface brightness,
they by no means constitute a statistical sample of any sort.  They
are not drawn randomly from the FGC, and thus their properties are not
representative of that catalog as a whole.  Because our selection of a
subsample was far from unbiased, the sample cannot be used for any
analysis of the numbers of galaxies of different surface brightnesses.
Likewise, the sample cannot be used to study the incidence of warping
at moderately bright surface brightness levels.

Finally, our subsample of the FGC includes galaxies with peak
$B$--band surface brightnesses (viewed edge-on) between 21.5 and
23$\surfb$ -- i.e.\ between the face-on value of the characteristic
Freeman (1970) surface brightness and the surface brightness limit of
the FGC survey data.  Thus, this limited range of $B-$band surface
brightness can become a selection effect which can influence some
results, such as apparent trends in color and extinction.  These
biases will be considered explicitly.


\section{Optical Observations, Data Reduction, \& Photometric Calibration}
\label{optobssec}

\subsection{Optical Imaging}

All optical observations of the FGC sample were made with the 2.5
meter Dupont telescope on Las Campanas, during the nights of September
22 and 23 in 1997, and March 29 and 30 in 1998, using a thinned
Tektronix 2048$^2$ CCD (``Tek5'') with 0.259$\arcsec$ pixels, a gain
of $\sim2.4$ DN$/e^{-}$, and readnoise of $\sim7e^{-}$.  The
conditions were photometric for the duration of all four nights.  For
each galaxy, a series of three exposures was taken through both a
Johnson $B$ and a Kron-Cousins $R$ filter, with the position of the
telescope shifted by more than 2$\arcmin$ between exposures to
minimize large and small scale flat-fielding variations, cosmic rays,
and cosmetic defects on the CCD.  Typical exposure times for the fall
subsample were 300 seconds per frame in $B$, and 120 seconds per frame
in $R$, for combined exposure times of 15 minutes and 6 minutes
respectively.  In order to maximize our chances of detecting extended
stellar halos in the nearby spring sample, our exposure times for the
spring sample were 2--3 times longer.

While the March 1998 observations were made during new moon, the
September 1997 observations were made with 40-50\% moon illumination
for the second half of each night.  During the night of September 23,
1997, time limitations forced us to image FGC 164, FGC 215, FGC
225, FGC 256, FGC 349, FGC 436, FGC 442 and FGC 446 while the moon was
up; as a result of the increased sky brightness, the $B$ images of
these galaxies are noticeably shallower.  Typical sky brightnesses
without moon were $22.2\pm0.2\surfb$ and $20.6\pm0.2\surfb$ in $B$ and
$R$ respectively, and $21.5\pm0.5\surfb$ and $19.9\pm0.5\surfb$ after
the moon had risen fully.  The total exposure times, sky brightnesses,
seeing FWHMs, and fluctuation levels of the reduced images are listed
in Table \ref{optobstable}.

\subsection{Image Reduction}

Each raw image has a bias level consisting of a time-variable mean
which changes by $\pm$ 1 DN between images and causes $\pm$0.5 DN
vertical structure along columns. There is also a stable component,
which is a small exponential decay over the first 150 pixels in every
row.  The time-variable bias level can be identified in the 30 columns
of overscan taken in each image. To remove it, we fit the average of
these columns with a 10th order polynomial and subtracted this fit
column by column.  A bias image was then made by averaging 15-20
overscan-subtracted, zero second exposures, with 2 iterations of $\pm
3\sigma$ rejection to remove cosmic rays. This image is then smoothed
with a 1x4 boxcar to reduce pixel-to-pixel noise to $\pm$ 0.3 DN, and
then subtracted from each image to remove the stable exponential
structure.  Dark frames show that the dark current in the CCD is less
than 0.5 DN per 900sec exposure with no two-dimensional structure
evident. No dark correction was therefore applied.

Pixel--to--pixel flat fields were generated separately for each night
using dome flats.  The resulting flat fields were divided into all
bias subtracted images, including twilight flats.  Except for
large--scale illumination changes due to the relative changes in the
telescope and dome positioning, differences between the dome flats on
adjacent nights were less than 0.3\%, peak--to--peak.  To enable faint
surface photometry over reasonably large areas (1--2'),
we have taken particular care with the large scale flat--field
calibration of our images. To remove large--scale illumination
features and to correct for the difference in color between the night
sky and the dome flats, twilight sky flats and night sky flats were
created.  The B-- and R--band twilight sky flats contained over
100,000 counts, cumulative.  

The night sky flats were made by
averaging the science exposures after rescaling them to a common
exposure time (multiplicative) and mean sky level (additive) and using
IRAF's CCDCLIP rejection with a grow radius of 10-30 pixels to
eliminate stars, galaxies, and cosmic rays; roughly 30 images per
night were used in this average. As the object galaxies were moved by
several arcminutes between exposures, the target objects were rejected
cleanly, and roughly 75\% of the images could usefully contribute to
the flat field in any region of the CCD. We then median smoothed the
combined images on a $\sim 2''$ scale to reduce the pixel--to--pixel
noise to $\sim0.001{\rm\,DN}$ ($\sim\!28\surfb$ for a sky level of
$20.5\surfb$), while preserving the large scale illumination
correction.  With the exception of one dust feature, the twilight and
night sky flats had peak--to--peak amplitude of much less than 0.5\%
(corresponding to roughly $26.5 R\surfb$ or $28.3 B\surfb$). 

For the moonless portion of the September 1997 observations, the
illumination was very uniform, and correction with twilight flats
alone yielded images in which no repeating large scale structures were
evident at the 0.5\% level between the flat-fielded science images.
Diffuse optical emission from galactic cirrus could clearly be
detected in some of the lower latitude fields, immediately suggesting
that flat-fielding residuals were small.  High frequency structure
(e.g. dust ``donuts'') in the $R$--band twilight flat was isolated by
dividing a heavily smoothed version (26$\arcsec$) of the flat into the
original.  This provided a high--frequency color correction to the
dome flats, the maximum amplitude of which was $\lesssim 1$\%.

Because of the color change of the sky after moonrise, separate night
sky flats were generated from and applied to the $B$-- and $R$--band
September 1997 observations taken after moonrise.
Night sky flats were also applied to all of the March 1998 observations,
for which the color and illumination of the twilights were not a good
match to the night sky during science observations.

The images were aligned by matching SExtractor (Bertin \& Arnouts
1996) positions for all objects in the $R$ and $B$ images. The
resulting pairs were used to shift the images into accurate alignment
with one of the $R$-band images, using IRAF's GEOMAP and GEOTRANS
packages.  All images were then averaged together with $\pm4\sigma$
rejection to produce the final images.  To simplify later analysis,
both $B$ and $R$ band images were rotated to orient the galaxy
horizontally. The required rotation was identified by measuring the
position of each galaxy at more than 4 locations along the disk in the
$R$-band image.  Finally, a 982x982 subsection was extracted from each
image, centered on the galaxy.

\subsection{Photometric Calibration}

Photometric calibration for the optical observations was
straightforward due to the nearly ideal conditions during the run.  At
least two separate Landolt (1983, 1992) standard fields were observed
at the beginning and end of the night, at a variety of exposure times
and at airmasses between 1.1 and 2.5. At least two more sets of
standard observations were made during the course of each night, giving
a total of 35-57 individual standard star measurements per
night through each filter.  The fluxes of the standard stars were
measured within the $14\arcsec$ diameter aperture used by Landolt.  For
the Fall run, the photometric solution was made assuming a constant
zero point and color term, with an airmass term allowed to vary
night-to-night.  For the Spring run, the zero point, color term, and
airmass were allowed to vary each night; the resulting
terms agreed to within $\pm1\sigma$.  Residuals from the best fit
photometric solution were typically $\sigma_m=0.012-0.020$.  The
resulting solutions are given in Table \ref{optphotsoltable}. 

Few of the galaxies in our sample have been extensively observed in
the optical.  As a result, we have
few means to check our calibration for consistency with other authors.
One galaxy from our sample, FGC E1371, was also studied in the ESO-LV
catalog (Lauberts \& Valentijn 1989; ESO-LV 3380010).  The reported
ESO-LV magnitudes are $B_{25}({\rm Cousins})=18.15\pm0.09$,
$B_{26}({\rm Cousins})=17.95\pm0.09$, $R_{25}({\rm
Cousins})=16.22\pm0.09$, and $R_{26}({\rm Cousins})=16.06\pm0.09$.  In
our filter system (Johnson $B$, Cousins $R$), we find $B_{25}({\rm
Johnson})=18.27$, $B_{26}({\rm Johnson})=18.04$, $R_{25}({\rm
Cousins})=16.23$, and $R_{26}({\rm Cousins})=16.21$, giving offsets of
+0.12, +0.09, $+0.01$, and $+0.15$ from the ESO-LV measurements.
These results are consistent within the errors ($\pm0.15$ for our sample).


\section{Infrared Observations, Data Reduction, \& Photometric Calibration} 
\label{irobssec}

\subsection{Infrared Imaging}

The infrared observations of the FGC sample were made with the duPont
2.5m telescope at Las Campanas Observatory, using an updated version
of the IRCAM camera originially described in Persson et al.\ (1992),
with the upgraded camera being similar to the P60IRC described by
Murphy et al.\ (1995).  The IRCAM, which consists of a Rockwell
NICMOS3 256x256 HgCdTe chip with $40\mu$ pixels, was operated at f/7.5
giving $0.348\arcsec$ pixels and an $89\arcsec$ field-of-view.  All
observations were made through the $K_s$ filter (developed by M.\
Skrutskie and described in the Appendix of Persson et al.\ 1998),
which cuts off at $\sim2.2\mu$ to reduce the thermal background by a
factor of two.  All observations took place during three runs: 21-22
September 1997, 10-14 October 1997, and 13-14 April 1998.  Observing
conditions on usable nights were as follows: September 21 \& 22,
photometric; October 10, clear for the first half of the night;
October 11, photometric with high winds and poor seeing; October 14,
photometric with very high winds and a 7.8 earthquake in the middle of
the night; April 13, non-photometric; and April 14, photometric.  The
electronics for the camera were replaced between the Fall 1997 runs
and the Spring 1998 run, changing the gain from 4.8 e-/adu to 7.5
e-/adu and requiring a new linearity correction.

Observations were made by looping the camera through successive sets
of 6 exposures of 20 seconds (2 minutes total) at various positions,
typically shifting the telescope by 1/2 of the camera field-of-view
($\sim35-45\arcsec$) between exposure loops.  Because of the thinness
of the FGC, our mosaic pattern allowed at least part of the galaxy to
be on the chip during every exposure, giving us high observing
efficiency without affecting our ability to create sky frames.  The
shifts were in the direction which maximized the overlap between the
chip and the galaxy (i.e. galaxies aligned East-West were dithered
North-South).  This strategy will limit our ability to reliably
interpret faint infrared structures more than $20\arcsec$ above the
planes of the galaxies.  However, given the existence of low-level
ghosting in the IRCAM and the faintness of the FGC subsample, such
scientific inquiries are beyond the limits of the data, regardless.

Total exposure times varied widely for our galaxies.  In keeping
with the wide range in $K_s$ band surface brightness, which spanned
4$\surfb$, the most massive galaxies were well exposed in only 12 minutes
of observations, while the lowest mass galaxies were still barely
detectable after several hours of integration.  Table \ref{irobstable}
lists the UT dates of all our observations, along with the exposure
times and estimates of the photometric quality (discussed in \S\ref{irredsec}).

\subsection{Image Reduction}    \label{irredsec}

Before processing, all images were linearized using scripts kindly
supplied by S. E. Persson.  (These scripts differed slightly between
1997 and 1998, due to the change in electronics.)  The linearization
was tested with data taken during the night of October 12 1997, while the
dome was closed due to bad weather.  For count levels less than
20,000 cts/pixel, the linearized data was linear to better than 0.1\%;
for reference, our typical sky levels in 20 second exposures were 6000
cts/pixel or less.  The linearization procedure was not tested during
the 1998 run.

All loops of exposures at a single position and exposure time were
averaged together using $\pm5\sigma$ rejection to remove cosmic rays.
Dark frames were created each night by combining loops of 50-100
exposures at every exposure time used for our science observations,
and were subtracted from the appropriate images.  On the night of
April 14 1998, darks were taken in both the evening and morning,
and were found to differ by $\sim15\%$; no explanation for this
variation was evident, and the two sets of darks were simply averaged
together.  The data from September 20 1997 were reduced using darks
taken the following night.  

Domeflats were derived most nights and divided into the summed,
dark-subtracted images.  Exceptions were the nights of October 10 and
11, for which domeflats taken during the night on October 12th were
used. The domeflats from October 10 and 11 were obtained during the
daytime with much warmer temperatures, and approached the level of
non-linearity.  Domeflats taken on the night of April 13 produced
large scale illumination residuals relative to twilight flats taken on
April 14. Domeflats taken on April 15 were used instead. (No domeflats
were taken on April 14 because the calibrations lamps broke.) 
After dark subtraction and flattening, known bad pixels were replaced
with locally interpolated values.  Henceforth, we will use ``images''
to refer to these coadded, calibrated frames, and not the individual
loop sub-exposures.

Sky subtraction and image alignment was performed with a modified
version of the DIMSUM V2.0 package.  First, a running sky image was
created from the median of every 6 adjacent sets of co-added images,
and then subtracted from the central image of the time series.  At the
beginning and end of each time series, no fewer than 4 adjacent images
were used to create the sky image.  Next, a rough offset was
calculated using the centroid of a single star in each image, and then
refined using IMALIGN with all stars available in the frame.  These
offsets were used to align the frames and then co-add them into a
single image.  The deep image was then used to create a mask of all
objects.  While the standard DIMSUM masking procedure works well for
compact, reasonably well-exposed objects, it was necessary to modify
DIMSUM's masking procedure to create appropriate masks for the
extended, low surface brightness galaxy in the frame.  The masks were
then de-registered, and used to create refined sky images for each
frame.  The new sky-subtracted images were then realigned and coadded
to produce the final image.  During the final alignment, the images
were expanded by a factor of 4 to avoid loss of resolution when
shifting the images.  For images of standard stars, a sky image was
created from all science exposures for the night and then scaled to the image
median and subtracted.  The resulting IR images were rotated and
aligned with the optical $R$ band images using IRAF's GEOMAP and
GEOTRANS packages.  To reduce uncertainties in aligning a single
image, after one pass through all of the data, the mean scale factor
was derived for the transformation between $K_s$ and $R$.  The images
were then realigned using the fixed scale factor.

While the seeing during our infrared run ($\sim1.1\arcsec$) was not
ideal, our final image resolution was also affected by the difficulty
in aligning some of our images.  Many of the galaxies were at
sufficiently high galactic latitude that there were no bright stars
visible in the individual images.  As the galaxies themselves were
often invisible in 120s exposures, and the duPont control software did
not record pointing position until 1999, some of the co-added images
were aligned using under-exposed faint stars, leading to non-spherical
PSFs in the co-added frames.  We were also troubled by high East-West
windshake, compounding this problem.  To account for this, we have fit
all stars in our $K_s$ images with elliptical Gaussian profiles.  The
PSF was measured for all available stars selected from $R$-band SExtractor
catalogs (CLASS\_STAR$\!>\!$0.9).  The mean properties of the PSF were
calculated using the brightest half of the stars (as measured in
$K_s$), iteratively rejecting outliers at the $\pm3\sigma$ level.  In
Table \ref{irobstable}, we report the resulting mean FWHM along the
major axis of the PSF ellipse, the axis ratio of the PSF ellipse, and
the position angle of the ellipse, measured relative to the position
angle of the galaxies.  Thus, galaxies whose $K_s$ PSFs have position
angles less than $\pm45^\circ$ will have worse seeing along the plane
of the galaxy, than perpendicular to the plane.  In general, due to
the paucity of bright stars in our images, the higher order shape
parameters for the PSF are not particularly well measured in most
cases (given that they are based upon 2-5 faint stars).  Values within
parentheses are based upon only a single star, and entries marked with
a "-" had no stellar objects within the frame.

For galaxies whose images were obtained over the course of 2 or more
nights, we reduced the images for each night separately.  The final image
was a weighted sum of the images from separate nights.

\subsection{Photometric Calibration}            \label{sec:irphot}

Calibration was performed using the faint standard star sequence
established by Persson et al.\ (1998).  Each standard star was moved
through the four quadrants of the chip to reproduce differences
between the four amplifiers.  Sets of standards were observed at the
beginning and end of each night, and usually 3-5 times during the
course of the night.  The fluxes of the standards were measured within
$10\arcsec$ diameter apertures, as in Persson et al.\ (1998).

For the 1997 data, a single magnitude zero point and color term was
found by simultaneously solving the four nights with high photometric
quality and allowing the extinction term for each night to vary separately.
The resulting photometric solutions and associated errors ($\sim0.02$
magnitudes) are given in Table \ref{irphotsoltable}.  Because of the
updated electronics between 1997 and 1998, the solution for the one
photometric night in 1998 was derived independently.

Even in non-photometric conditions (i.e. any cloud cover visible in
any part of the sky), standards were monitored to test our ability to
judge the level of cloud cover.  The derived ``photometric'' solutions
for the parts of non-photometric nights which were judged to be
reasonably clear are listed in parentheses in Table
\ref{irphotsoltable}.  In general, the solutions are nearly identical
to those derived for the photometric nights, suggesting that the parts
of the night which we considered to be clear had transparencies
comparable to truly photometric conditions.  As a further test, we
made plots of the sky level as a function of time during our science
exposures.  On the photometric nights, the sky level was very stable,
varying extremely slowly and smoothly.  By comparing our observing
logs to plots of the sky level during partially cloudy nights, we
found that the presence of clouds produced obvious, rapid variations
in the sky level.  We used the combination of our observing notes and
plots of the sky level to judge the degree of photometric accuracy
reported in Table \ref{irobstable}.

Unlike the $B$ and $R$--band results, we were unable to find any
infrared observations of our sample galaxies in the literature with
which to compare our global calibration.  While roughly 1/4 - 1/3 of
our sample is visible in the existing 2MASS survey (Skrutskie et al.\ 1997),
these galaxies are so faint and diffuse that they have not been
cataloged or photometered.

We have, however, checked our photometric calibration internally.  For
roughly half of our sample, we obtained images over two or more
nights.  This was particularly true for the galaxies with the faintest
$K_s$-band surface brightnesses, which required much deeper
observations.  We have performed aperture photometry on all fields
observed on separate nights in order to (1) test the consistency of
our photometry and (2) to correct observations taken in
non-photometric conditions.  The mean amplitude of night-to-night
variations among the photometric observations and the ``questionable''
observations (marked by a ``:'' in Table \ref{irobstable}) were
identical ($\Delta |m|\!=\!0.04$), and were all within the $2\sigma$
uncertainties defined by photon counting and the photometric
calibration.  There were also no consistent offsets between nights
(i.e.  the offsets were just as likely to be negative as positive).
Only the non-photometric observations marked with double colons in
Table \ref{irobstable} showed significant variations, with $\Delta
m_{\rm FGC 143}\!=\!0.09$, $\Delta m_{\rm FGC 256}\!=\!0.31$, $\Delta
m_{\rm FGC 901}\!=\!0.55$, and $\Delta m_{\rm FGC 1971}\!=\!0.28$.
The zero points of these non-photometric images were adjusted
accordingly before they were coadded with the photometric data.  We
also did not include the data for FGC 1303 from 980415 and for FGC 277
from 971015, as unfortunately, the sub-images could not be properly
aligned.

\section{Masking}              \label{masksec}

In order to facilitate analysis of the galaxy profiles at low light
levels, we generated a mask for each image, to identify regions
contaminated by interloping sources (stars, galaxies, meteor trails,
etc.).  The masks were made using our $R$ band images, which typically
had the highest signal-to-noise.  First, SExtractor (Bertin \& Arnouts
1996) was used to identify all objects in the frame.  Using the
reported ellipticities and isophotal areas, we masked elliptical
regions around all detected objects (except for the immediate vicinity
of the central galaxy where objects were masked by hand to avoid
masking HII regions associated with the galaxy).  We increased the size
of the masked regions by a factor of three, producing a factor of 9
increase in the area, and thus reducing any contamination from
interloper objects at low light levels.

The resulting masks were then visually inspected, and edited by hand
to remove any remaining sources of contamination in all three bands
(i.e.\ $B$, $R$, \& $K_s$).  Examples of these include objects which
fell on top of or very near to a galaxy (a problem in lower latitude
fields), meteor trails, diffraction spikes from bright stars, poorly
removed bad columns, obvious scattered light from stars just off the
field, and regions which were not covered in all three sub-exposures
used to make the final optical images.  There is some ambiguity about
removing faint ``contaminating'' objects close to the galaxy, given
the natural confusion between faint intervening galaxies and small HII
regions within the galaxy.  This confusion is highest in the furthest
outskirts of the galaxies, which are known to harbor faint HII
regions, in spite of having little diffuse optical emission (e.g.\ 
Ferguson et al.\ 1998).  We typically mask
these regions when there was no diffuse emission connecting
the faint object to the main galaxy.  Given that these sources are all
extremely faint, their inclusion or exclusion will make little
difference in the total magnitude of the galaxy.  They may slightly
affect the shapes of the faintest isophotes, however.

We also generated a second set of masks to generously encompass the
faintest possible isophotes of the main FGC galaxy.  These masks are
used to exclude all possible contributions from the galaxy when
measuring the background sky.

Finally, for each galaxy, identical masks were used for all bands,
so that all analysis was restricted to the
same portions of the images.

\section{Sky Subtraction}          \label{skysubsec}

After the $B$ and $R$ images were flattened using the combination of
dome flats, twilight flats, and supersky flats, there remained
residual diffuse low surface brightness structure visible in many of
the images.  These structures varied significantly from image to
image, and are thus are not due to variations in the illumination or
the response of the CCD.  In most cases, the position of the structure
shifts with the sky in the series of dithered images, suggesting these
sky fluctuations are an astronomical source and not a calibration
problem. These remaining variations are most likely the result of a
combination of scattered light from stars beyond the field of view,
and optical emission from the 100$\mu$ cirrus (e.g.\ Guhathakurta et
al.\ 1990).  These unavoidable sources of non-uniform background
represent a fundamental limit on our ability to trace the structure of
galaxies to extremely low surface brightnesses ($\mu(R)>28\surfb$).

Because these sources are additive, we have made a first-order attempt
to subtract them from the $B$ and $R$ images by fitting a plane to the
982$\times$982 pixel subregion around each galaxy.  The images were
masked with both the background object mask and the galaxy mask
(\S\ref{masksec}), leaving only sky pixels in the resulting image. A
plane was then fit to the unmasked background, using iterative
rejection of outliers.  The resulting slopes implied a variation
across the $4.2\arcmin$ region of typically fainter than
$\Delta\mu>\!29\,B\surfb$ and $\Delta\mu>\!27\,R\surfb$.  These
surface brightness variations are also characteristic of the smaller
scale structures found in some of the images.

We chose not to perform a further background subtraction on the $K_s$
band images.  Because of the smaller field of view of the infrared
images and our generous masking of the galaxy region, the number of
unmasked ``background'' pixels was small, leading to unrepresentative
fits to the overall background.  The unmasked pixels were
typically found in the outskirts of the image, which were sampled by a
much smaller number of sub-images, and thus were of lower signal-to-noise
and more prone to statistical variations in the sky level.
Furthermore, given the higher level of ghosting and scattered light in
the IRCAM (compared to the optical CCD camera) and the brighter IR
sky, remaining sky variations are more likely to be calibration
problems than true astronomical signals.  Experiments quickly revealed
that fitting the sky background with the same procedure used for the
optical bands created gradients across the central field, rather than
eliminating them.


\section{Isophotes and Integrated Magnitudes}          \label{magsec}

Before defining magnitudes for the galaxies in our sample, it was
necessary to define isophotal contours for photometry.  Using IDL,
we defined contour levels in the masked, unsmoothed image, down to an
isophote level of 3$\sigma$ above the sky.  To trace the contours to
fainter limits, we developed a variable smoothing method, wherein the
image was smoothed with a Gaussian ellipse whose size was adjusted to
preserve constant signal-to-noise in the resulting image, with the
maximum smoothing length fixed to a width of 15 pixels (3.9$\arcsec$)
and a 3:1 axial ratio oriented along the major axis of the galaxy.  
Note, however, that while smoothing was used to set the shape of the
faintest isophotes, photometry was performed only on the unsmoothed
images.  We also did not correct the levels of the isophotes to
compensate for foreground Galactic extinction; the isophotes
therefore refer to an apparent surface brightness, rather than one
intrinsic to the galaxy.

The resulting isophotes are plotted in Figure \ref{contourfig}.  The
plots clearly show the transition between using smoothed and
unsmoothed isophotes, as individual pixels mark the edge of the
latter, and smoother contiguous lines mark the former.  Occasionally,
contours skirt the perimeters of masked stars, leading to odd shapes
in the isophotes.  To partially compensate for the flux lost to
masking, and to better trace the isophotes, we derived approximate
smooth models for the galaxies in an attempt to ``fill'' in the masked
regions.  As this process is necessarily uncertain, we include the
magnitude of the flux added by the model in our uncertainties; rarely
is this contribution the dominant source of error.

Using these isophotes, we have derived isophotal magnitudes in $B$,
$R$, and $K_s$. We
present these in Table \ref{magtable} for our sample, uncorrected for
galactic extinction.  We give magnitudes in reference to a specific
isophotal level in each band ($\mu_{lim}(B)\!=\!27\surfb$,
$\mu_{lim}(R)\!=\!26\surfb$, and $\mu_{lim}(K)\!=\!22\surfb$), in
order to facilitate comparison with other comparable
observations, most notably the Ursa Major cluster sample of Verheijen
(1997).  Because the isophotal areas associated with these limits are
a strong function of the filter bandpass, we also give magnitudes in
$B$ and $K_s$ which are calculated within the area of the
$\mu_{lim}(R)\!=\!25\surfb$ isophote; these magnitudes should be used
when determining colors for the sample.  In a few cases, due to
somewhat shallower observations or larger problems with scattered
light (e.g.\ FGC 2292), the reference isophotes were not reliably
determined; these cases are left blank in Table \ref{magtable}.
Because projection makes the galaxies in this edge-on sample appear
brighter, the isophotes of our sample probably occur at larger radii
than they would if seen face-on, and are thus comparable to
fainter isophotes for less inclined galaxies.  We have chosen not to
correct the final magnitudes listed in Table \ref{magtable} to either
total or face-on values, as such conversions would be highly uncertain.

In calculating the magnitudes, we have included the color term in the
zero point by first calculating the magnitudes assuming a mean color
of $B-R\!\sim\!1$ and $J-K_s\!\sim\!1$, and then making a second order
correction (typically less than 0.02) based upon the resulting color
of the galaxy within the $\mu_{lim}(R)\!=\!25\surfb$ isophote.  As we
did not have information on the $J-K_s$ colors of the sample, we made
no further correction to the infrared $K_s$ magnitudes.

Because the amplitude of Galactic extinction corrections tend to be a
function of time and a matter of taste, we have listed $E(B-V)$ from
Schlegel et al.\ (1998) in Table \ref{magtable}, but have not include
the associated correction in the $B$, $R$, or $K_s$ values in Table
\ref{magtable}.

\subsection{Uncertainties \& Reliability of Faint Isophotes}

There are three main sources which contributed to the uncertainties
given in Table \ref{magtable}.  First are the photometric calibration 
uncertainties, $\sigma_m$, given
in Tables \ref{optphotsoltable} \& \ref{irphotsoltable}, which are
typically $\lesssim0.02$ in the optical and $\lesssim0.05$ in the infrared.
The value of $\sigma_m$ is the RMS scatter around the photometric
solution, and is an empirical measurement of the characteristic error
in an individual measurement.  Thus, $\sigma_m$ is a conservative
estimate of the photometric uncertainty associated with any single
observation.

The second source of uncertainty comes from the area lost to masking
of stars on or near the galaxy, or to the small area of the $K_s$
image (smaller than the extent of the $R$-band isophotes in some
cases).  Both effects cause flux to be underestimated.
In most instances, the galaxies in our sample are at high enough
galactic latitude that few foreground stars overlap even the faintest
isophotes of the galaxy, and thus this is usually not a problem.
However, we have attempted to correct for this effect, by using the
smoothed models described above to interpolate within the masked
regions.  While this process is uncertain, it is at least a step in
the right direction.  Furthermore, the magnitude of the correction
gives some indication of the magnitude of the uncertainty.  Typically,
these corrections are less than 0.05 magnitudes.

The final, most subtle, and dominant source of uncertainty comes from
sky subtraction.  If the level of the sky is wrong, then this
contributes/removes additional light to/from the galaxy, in proportion
to the area of the isophote used for photometry.  To set the
uncertainty in the sky subtraction, we have analyzed the noise
properties of the unmasked ``sky'' pixels (see \S\ref{skysubsec}) for
a variety of boxcar smoothing lengths, $L$  (Figure \ref{noisefig}).
For an uncorrelated, uniform sky background, the standard deviation of
the sky pixels, $\sigma_{\rm sky}$, will decrease with increasing
smoothing lengths as $1/L$; this relation is plotted as the dashed
line in Figure \ref{noisefig}.  For a highly correlated, non-uniform
sky background, $\sigma_{\rm sky}$ will be constant as long as $L$ is
smaller than scale of the non-uniformity, and the Poisson fluctuations
in the smoothed image are smaller than the amplitude of the
non-uniformity (for example, imagine a step-function in brightness
across the image).
Mathematically, for a given power spectrum of background sky
fluctuations, the predicted distribution of $\sigma_{\rm sky}(L)$ can
be calculated in analogy to the ``counts-in-cells'' formalism
developed for large-scale structure analysis (e.g.\ Peebles 1980).
The inclusion of the correlated background increases the variance
above that predicted for pure Poisson fluctuations by adding a term
involving the integral of the correlation function over the smoothing
area.
In each image, we can therefore asses the level of residual structure
in the sky and its origin from the behavior of $\sigma_{\rm sky}(L)$,
shown in Figure \ref{noisefig}. These plots also indicates the
reliability of surface photometry at various length scales.

The plots of $\sigma_{sky}(L)$ in Figure \ref{noisefig} reveal a
number of facts about our data.  First, there is indeed correlated
structure in the sky, as revealed by the deviation of the measured
curves from the $1/L$ behavior expected for a uniform Poisson
background.  The presence of such structures is not too surprising,
given that we fully expect to have residual contamination from
scattered light and from stars and galaxies which lie below our
detection threshold.  

Second, the similarity of the $B$ and $R$ band $\sigma_{sky}(L)$
curves, down to the faintest surface brightnesses, suggests that the
deviation from the $1/L$ Poisson expectation does not result from
uncertainties in flat-fielding, at least for the optical imaging.
Large-scale flat-fielding errors would be expected to differ between
the two bands, leading to shape variations for observations taken
through different filters.  These shape variations would be
consistent, however, for all galaxies observed on a single night.
Although we do see occasional variations in between the shape of the
$B$ and $R$ curves (c.f.\ FGC 51), these variations are not consistent
with other galaxies observed on the same night.  Thus, they are more
realistically interpreted as being variations in the scattered light
in the different bands (see below), given that the pointing centers
are not identical for the sub-images used to make the $B$ and $R$
images.

Third, we also see many cases where the curve is rolling over to flat
as we approach the amplitude of the background sky variations.  In
some cases (e.g.\ FGC 227, FGC 2548, FGC E1371, FGC E1404, \& FGC
E1440), optical emission from 100$\mu$ cirrus is clearly visible in
the frame.  In others (c.f.\ FGC 51, FGC 2264, \& FGC 1642), scattered light
is a large problem in one of the sub-images.  In all of these cases,
$\sigma_{sky}(L)$ rolls over at large smoothing lengths, as expected,
and reveals the characteristic brightness of the structure.

Finally, the plots in Figure \ref{noisefig} clearly demonstrate the
well known difficulties with attempting to do reliable infrared
surface photometry of nearby galaxies.  The small field-of-view of IR
detectors, the brightness and variability of the IR sky, and the
difficulty in constructing reasonable dome flats all conspire to make
the prospect of accurate faint surface brightnesses photometry
daunting, if not practically impossible with current detectors.  The
$K_s$-band $\sigma_{sky}(L)$ curves deviate in shape from the optical
curves at relatively small smoothing lengths ($\sim\!5\arcsec$) and
bright surface brightnesses ($\mu_{K_s}\!\lesssim\!22\surfb$), demonstrating
the limitations of our $K_s$-band data at large scales and faint surface
brightnesses.  Because of mosaicking, the signal-to-noise
of the $K_s$ images tends to degrade towards the outskirts of the image,
away from the galaxy.  The estimates of the sky uncertainty come from
these outer regions, and will thus be biased towards higher values.
Thus, sky subtraction near the galaxy is therefore likely to
be somewhat better than indicated by Figure \ref{noisefig}.

To treat the contribution that the uncertainties in sky subtraction
make to the magnitudes in Table \ref{magtable}, we take
the faintest, reliably--determined value of $\sigma_{sky}(L)$ (i.e.\ the
highest point in each curve plotted in Figure \ref{noisefig}) as being
characteristic of the error in our determination of the sky.  We then
calculate the uncertainty in the total flux over the isophotal area.

\section{Summary}                  \label{summarysec}

In this first of a series of papers, we have described our sample of
edge-on, late-type disk galaxies.  As we have demonstrated, the
optical and infrared imaging data on these galaxies is exceptionally
well characterized, and will be well suited for upcoming analysis of
the structural properties of the sample, down to very faint surface
brightness limits ($\mu(B)\sim29.5\surfb$, $\mu(R)\sim29\surfb$).


\bigskip
\bigskip
\centerline{Acknowledgements}
\medskip

First and foremost, we are deeply indebted to the generous allocations
of telescope time made by Carnegie Observatories during the course of
this project.  Likewise, the success of this work was dependent upon
the excellent instrumentation and telescopes maintained by the staff
at Carnegie and at Las Campanas, and its execution would not have been
nearly as smooth without the expert help of Herman Olivares and
Fernando Peralta.  We are also grateful to the hospitality of Casa
Polacka during times of inclement weather, and to Scott Trager for his
company during the near fatal empanada excursion during the October
1997 observing run.  Eric Deutsch is warmly thanked for help with IDL,
as is Dan Rosenthal for spirited discussions.

JJD was partially supported through NSF grant AST-990862, and
submitted this paper while at the Institute for Theoretical Physics at
UC Santa Barbara, which is supported in part by the National Science
Foundation under Grant No.\ PHY94-07194.  Support for RAB was provided
by NASA through Hubble Fellowship grant HF-01088.01-97A awarded by
STScI, which is operated by AURA, Inc. for NASA under contract NAS
5-2655.  This research has made use of the NASA/IPAC Extragalactic
Database (NED) which is operated by the Jet Propulsion Laboratory,
California Institute of Technology, under contract with the National
Aeronautics and Space Administration.  This paper also made use of the
Digitized Sky Surveys, which were produced at the Space Telescope
Science Institute under U.S.  Government grant NAG W-2166. The images
of these surveys are based on photographic data obtained using the
Oschin Schmidt Telescope on Palomar Mountain and the UK Schmidt
Telescope. The plates were processed into the present compressed
digital form with the permission of these institutions. The National
Geographic Society - Palomar Observatory Sky Atlas (POSS-I) was made
by the California Institute of Technology with grants from the
National Geographic Society.

\vfill
\clearpage

\section{References}

\hi{Bahcall, J. N. 1984, \apj, 276, 156.}

\hi{Bertin, E., \& Arnouts, S. 1996, \aasup, 117, 393.}

\hi{de Blok, E., \& van der Hulst, J. M. 1998, \aanda, 336, 49.}

\hi{de Blok, E., \& McGaugh, S. S. 1997, \mnras, 290, 533.}

\hi{de Grijs, R., 1998, \mnras, 299, 595.}

\hi{Dove, J. B., \& Thronson, H. A. 1993, \apj, 411, 632.}

\hi{Ferguson, A. M. N., Wyse, R. F. G., Gallagher, J. S., \& Hunter, D. A. 1998, \apjl, 506, 19.}

\hi{Freeman, K. C. 1970, \apj, 160, 811.}

\hi{Goad, J. W., \& Roberts, M. S. 1981, \apj, 250, 79.}

\hi{Giovanelli, R., Avera, E. \& Karachentsev, I. D. 1997, \aj, 114, 122.}

\hi{Guhathakurta, P., Tyson, J. A., \& Majewski, S. R. 1990, \apjl, 357, 9.}

\hi{Haynes, M. P., Giovanelli, R., Herter, T., Vogt, N. P., Freudling, W., Maia, M. A. G., Salzer, J. J., \& Wegner, G. 1997, \aj, 113, 1197}

\hi{Landolt, A. U. 1983, \aj, 88, 439.}

\hi{Landolt, A. U. 1992, \aj, 104, 340.}

\hi{Lauberts, A., \& Valentijn, E. A. 1989, The Surface Photometry
        Catalogue of the ESO-Uppsala Galaxies (ESO-LV), ESO.}

\hi{Karachentsev, I. D., Karachentseva, V. E., \& Parnovsky, S. L.\ 1993,
Astro. Nacht., 314, 97.}

\hi{Kudrya, Y. N., Karachentsev, I. D., Karachentseva, V. E., Parnovskii, S. L. 1994, Astr. Letters., 20, 8.}

\hi{McGaugh, S. S. 1994, \apj, 426, 135.}

\hi{Moshir et al 1990, IRAS Faint Source Catalog.}

\hi{Murphy, D. C., Persson, S. E., Pahre, M. A., Sivaramakrishnan, A., \&
Djorgovski, S. G. 1995, \pasp, 107, 1234.}

\hi{Olling, R. P. 1995, \aj, 110, 591.}

\hi{Persson, S. E., Murphy, D. C., Krzeminski, W., Roth, M., \& Rieke, M. J.\
1998, \aj, 116, 2475.}

\hi{Persson, S. E., West, S. C., Carr, D. M., Sivaramakrishnan, A., \&
Murphy, D. C.\ 1992, \pasp, 104, 204.}

\hi{Schlegel, D. J., Finkbeiner, D. P., \& Davis, M. 1998, \apj, 500. 525.}

\hi{Schneider, S. E., Thuan, T. X., Magri, C., \& Wadiak, J. E. 1990, \apjs, 72, 245.}

\hi{Skrutskie, M.F., Schneider, S.E., Stiening, R., Strom, S.E., Weinberg, M.D.,
Beichman, C., Chester, T., Cutri, R., Lonsdale, C., Elias, J., Elston,
R., Capps, R., Carpenter, J., Huchra, J., Liebert, J., Monet, D.,
Price, S. and Seitzer, P. 1997, in "The Impact of Large Scale Near-IR
Sky Surveys," p187-195, F. Garzon \et eds. (Dordrecht: Kluwer).}

\hi{Spitzer, L., 1942 \apj, 95, 329.}

\hi{Swaters, R. A., Madore, B. F., \& Trewhella, M. 2000, \apjl, 531, 107L.}

\hi{van den Bosch, F. C., Robertson, B. E., Dalcanton, J. J., \& de Blok, W. J. G. 2000, \aj, 119, 1579.}

\hi{van der Kruit, P. C., \& Searle L. 1982, \aanda, 110, 61.}

\hi{van der Kruit, P. C., \& de Grijs, R. 1999, \aanda, 352, 129.}

\hi{Verheijen, M. A. W. 1997, PhD Thesis, University of Groningen.}

\hi{Zasov, A. V., Makarov, D. I., \& Mikhailova, E. A. 1991, Sov.\ Astron.\
        Lett., 17, 5.}

\hi{Zwaan, M. A., van der Hulst, J. M., de Blok, W. J. G., \& McGaugh, S. S
1995, \mnras, 273, L35.}

\vfill
\clearpage


\pagestyle{empty}
\begin{deluxetable}{rrrrrrr}                    
\tablecaption{Survey Galaxies\label{positiontable}}
\scriptsize
\tablehead{
\colhead{FGC}      &    \colhead{UGC} &
\colhead{$\alpha$}      &   \colhead{$\delta$}      &
\colhead{PA}      &
\colhead{$V_\odot$} &   \colhead{$W_{50,c}$}
\nl
\colhead{}      &       \colhead{} &
\colhead{(1950.0)}      &   \colhead{(1950.0)}      &
\colhead{$(\deg)$}      &
\colhead{$\kms$} &      \colhead{$\kms$}
}
\startdata
31    &         & 00:17:00.2 & $+$18:22:53 &  34 &      5287  &  143  \nl
36    &         & 00:19:34.0 & $+$10:06:15 & 149 &      5446  &  191  \nl
51    &   290   & 00:26:32.3 & $+$15:37:32 & 133 &       767  &   69  \nl
84    &         & 00:42:07.2 & $-$11:27:43 &  48 &         -  &    -  \nl
130   &         & 01:08:17.0 & $+$14:01:04 &  22 &         -  &    -  \nl
143   &   819   & 01:13:24.0 & $+$06:22:50 &  74 &      2417  &   97  \nl
164   &   971   & 01:22:14.7 & $+$00:46:27 & 148 &      4735  &  119  \nl
215   &  1417   & 01:53:58.4 & $+$17:28:01 & 153 &     11430  &  300  \nl
225   &  1484   & 01:57:41.0 & $+$15:43:23 & 106 &      5063  &  172  \nl
227   &         & 01:58:12.5 & $+$19:27:50 &   9 &      5620  &  214  \nl
256   &  1677   & 02:08:26.0 & $+$06:25:56 & 135 &      1608  &   77  \nl
277   &         & 02:17:02.4 & $+$18:45:05 & 136 &         -  &    -  \nl
310   &         & 02:29:56.5 & $+$15:29:58 &  78 &      5873  &  202  \nl
349   &         & 02:48:43.3 & $+$05:21:02 & 173 &      8114  &  212  \nl
395   &  2548   & 03:09:10.1 & $+$00:51:29 & 134 &         -  &    -  \nl
436   &         & 03:32:00.9 & $+$14:58:25 & 169 &      6218  &  243  \nl
442   &         & 03:37:57.7 & $+$03:22:33 & 136 &      5733  &  205  \nl
446   &  2852   & 03:42:16.0 & $+$05:44:54 & 119 &      6101  &  334  \nl
780   &  4524   & 08:37:35.5 & $+$05:48:43 &  50 &      1939  &  150  \nl
901   &         & 09:29:17.5 & $+$12:28:57 & 170 &      5899  &  201  \nl
913   &         & 09:33:46.6 & $+$15:46:21 & 175 &      4329  &  154  \nl
979   &  5347   & 09:54:40.5 & $+$04:45:53 &  18 &      2156  &  210  \nl
1043  &  5537   & 10:13:04.2 & $+$07:34:33 & 146 &      3755  &  288  \nl
1063  &         & 10:21:55.7 & $+$12:09:56 & 168 &      2431  &  140  \nl
1285  &  6594   & 11:35:03.3 & $+$16:50:00 & 134 &      1040  &  150  \nl
1303  &         & 11:43:32.9 & $+$13:09:24 & 115 &      3290  &  134  \nl
1415  &  7394   & 12:17:54.0 & $+$01:44:40 & 146 &      1598  &  173  \nl
1440  &  7607   & 12:26:19.4 & $+$04:34:02 &  53 &      4240  &  301  \nl
1642  &         & 13:33:32.4 & $+$08:26:33 & 139 &      1243  &  110  \nl
1863  &  9760   & 15:09:30.3 & $+$01:53:11 &  55 &      2023  &  137  \nl
1945  & 10000   & 15:42:18.8 & $+$04:06:54 & 124 &      3541  &  206  \nl
1948  & 10025   & 15:43:53.0 & $+$03:00:00 &  81 &      1522  &  109  \nl
1971  & 10111   & 15:55:53.4 & $+$13:18:41 &  37 &     10387  &  454  \nl
2131  & 10852   & 17:23:56.6 & $+$11:21:35 & 163 &      2781  &  139  \nl
2135  &         & 17:25:06.6 & $+$13:42:17 & 124 &      9059  &  222  \nl
2217  & 11301   & 18:35:42.2 & $+$17:29:22 & 110 &      4500  &  478  \nl
2264  &         & 19:47:00.6 & $-$10:54:02 & 127 &         -  &    -  \nl
2292  &         & 20:52:36.3 & $+$17:28:02 & 170 &      5563  &  174  \nl
2367  &         & 22:05:34.8 & $+$15:28:14 &  25 &      7827  &  159  \nl
2369  &         & 22:07:21.6 & $+$07:11:01 &  83 &         -  &    -  \nl
2548  &         & 23:49:47.6 & $+$07:41:54 & 113 &      3865  &  142  \nl
2558  &         & 23:52:46.0 & $+$03:32:38 &  58 &      5378  &  178  \nl
E1371 &         & 19:18:33.6 & $-$38:18:07 & 141 &         -  &    -  \nl
E1404 &         & 19:46:40.1 & $-$36:30:16 &  35 &         -  &    -  \nl
E1440 &         & 20:08:56.0 & $-$22:46:12 &  10 &         -  &    -  \nl
E1447 &         & 20:12:16.6 & $-$55:13:25 &  23 &         -  &    -  \nl
E1498 &         & 20:35:21.0 & $-$53:26:31 &  64 &         -  &    -  \nl
E1619 &         & 21:32:44.7 & $-$28:55:56 & 149 &         -  &    -  \nl
E1623 &         & 21:33:58.0 & $-$19:49:30 &  12 &         -  &    -  \nl
\enddata
\end{deluxetable}

\clearpage

\pagestyle{empty}
\begin{deluxetable}{lcccccccc}          
\scriptsize
\tablecaption{Optical Observations\label{optobstable}}
\tablehead{
\colhead{FGC}      &
\colhead{Exp ($B$)}      &
\colhead{Exp ($R$)}      &
\colhead{$\mu_{sky}$ ($B$)}  &
\colhead{$\sigma_\mu$($B$)}  &
\colhead{$\mu_{sky}$ ($R$)}  &
\colhead{$\sigma_\mu$($R$)}  &
\colhead{FWHM ($B$)}          &
\colhead{FWHM ($R$)} \nl
\colhead{}      &
\colhead{$s$}      &
\colhead{$s$}      &
\colhead{$\surfb$}  &
\colhead{rms per $\Box\arcsec$}  &
\colhead{$\surfb$}  &
\colhead{rms per $\Box\arcsec$}  &
\colhead{$^{\prime\prime}$}          &
\colhead{$^{\prime\prime}$} \nl
}
\startdata
 31   & \phn900 &360 & 22.6 & 27.6 & 20.7 & 26.5 &1.0 & 0.9 \nl
 36   & \phn900 &600 & 22.6 & 27.5 & 20.6 & 26.8 &1.1 & 0.8 \nl
 51   & \phn900 &960 & 22.3 & 27.0 & 20.4 & 26.7 &1.1 & 0.8 \nl
 84   & \phn900 &480 & 21.9 & 27.3 & 20.8 & 26.6 &0.9 & 0.7 \nl
 130  & \phn900 &360 & 22.5 & 27.3 & 20.4 & 26.2 &1.0 & 1.0 \nl
 143  & \phn900 &240 & 22.5 & 27.4 & 20.4 & 26.3 &1.0 & 0.9 \nl
 164  & \phn900 &360 & 21.4 & 27.0 & 20.4 & 26.3 &1.1 & 1.0 \nl
 215  & \phn900 &360 & 21.0 & 26.8 & 20.2 & 26.1 &1.4 & 1.1 \nl
 225  & \phn900 &360 & 21.5 & 27.0 & 20.2 & 26.3 &1.1 & 0.9 \nl
 227  & \phn900 &360 & 22.4 & 27.3 & 20.1 & 26.2 &1.3 & 1.1 \nl
 256  & \phn900 &600 & 21.5 & 27.1 & 20.3 & 26.4 &1.1 & 1.0 \nl
 277  & \phn900 &600 & 22.4 & 27.4 & 20.1 & 26.3 &1.1 & 1.0 \nl
 310  & \phn900 &360 & 22.1 & 26.9 & 20.1 & 26.1 &1.2 & 0.9 \nl
 349  & \phn900 &360 & 21.4 & 27.0 & 20.3 & 26.3 &1.1 & 0.8 \nl
 395  & \phn900 &360 & 21.9 & 26.9 & 20.3 & 26.3 &1.0 & 0.8 \nl
 436  & \phn300 &240 & 20.6 & 25.9 & 19.7 & 25.8 &1.3 & 1.0 \nl
 442  & \phn900 &240 & 21.2 & 26.8 & 20.2 & 25.9 &1.3 & 0.8 \nl
 446  & \phn900 &360 & 21.1 & 26.7 & 20.2 & 26.2 &1.5 & 0.9 \nl
 780  &    2100 &900 & 22.3 & 27.7 & 20.4 & 26.8 &0.7 & 0.7 \nl
 901  &    1800 &600 & 22.1 & 27.8 & 20.5 & 26.7 &1.2 & 0.9 \nl
 913  &    1800 &900 & 22.2 & 27.7 & 20.3 & 26.9 &0.8 & 0.7 \nl
 979  &    1800 &600 & 22.4 & 27.7 & 20.6 & 26.6 &0.8 & 0.7 \nl
 1043 & \phn900 &360 & 22.3 & 27.3 & 20.6 & 26.4 &0.9 & 0.9 \nl
 1063 & \phn900 &720 & 22.3 & 27.3 & 20.5 & 26.9 &1.0 & 0.7 \nl
 1285 &    1800 &900 & 22.3 & 27.7 & 20.6 & 27.0 &1.0 & 1.1 \nl
 1303 &    2700 &900 & 22.2 & 28.0 & 20.7 & 26.9 &0.9 & 0.8 \nl
 1415 & \phn900 &360 & 22.2 & 27.2 & 20.7 & 26.6 &0.9 & 0.8 \nl
 1440 &    1800 &900 & 22.3 & 27.8 & 20.7 & 27.0 &0.8 & 0.7 \nl
 1642 &    2700 &600 & 22.3 & 27.8 & 20.7 & 26.7 &1.0 & 0.7 \nl
 1863 &    2700 &900 & 22.4 & 28.1 & 20.8 & 27.0 &0.8 & 0.7 \nl
 1945 &    1800 &600 & 22.4 & 27.9 & 20.8 & 26.8 &0.8 & 0.8 \nl
 1948 &    2700 &900 & 22.4 & 28.0 & 20.7 & 26.8 &1.0 & 0.8 \nl
 1971 &    2700 &900 & 22.4 & 28.1 & 20.7 & 26.9 &0.9 & 0.9 \nl
 2131 &    2700 &900 & 22.3 & 27.8 & 20.6 & 26.7 &1.3 & 1.0 \nl
 2135 &    1800 &600 & 22.3 & 27.9 & 20.7 & 26.7 &0.9 & 0.8 \nl
 2217 & \phn900 &360 & 22.1 & 26.5 & 20.5 & 25.7 &1.3 & 1.3 \nl
 2264 & \phn900 &360 & 22.4 & 26.2 & 20.8 & 25.7 &0.8 & 0.7 \nl
 2292 & \phn900 &360 & 22.6 & 27.3 & 20.7 & 26.2 &0.9 & 0.8 \nl
 2367 & \phn900 &360 & 22.7 & 27.3 & 20.7 & 26.3 &1.0 & 0.9 \nl
 2369 & \phn900 &360 & 22.7 & 27.5 & 20.9 & 26.4 &1.1 & 0.8 \nl
 2548 & \phn900 &360 & 22.5 & 27.5 & 20.9 & 26.7 &0.9 & 0.8 \nl
 2558 & \phn900 &360 & 22.5 & 27.4 & 20.8 & 26.4 &1.0 & 0.8 \nl
 E1371 &\phn900 &360 & 22.4 & 27.5 & 20.7 & 26.3 &0.9 & 0.9 \nl
 E1404 &\phn900 &360 & 22.4 & 27.4 & 20.7 & 26.2 &1.2 & 0.9 \nl
 E1440 &\phn900 &360 & 22.4 & 27.3 & 20.6 & 26.3 &0.9 & 0.8 \nl
 E1447 &\phn900 &360 & 22.7 & 27.6 & 21.0 & 26.6 &0.8 & 0.8 \nl
 E1498 &\phn900 &360 & 22.8 & 27.6 & 21.0 & 26.4 &0.9 & 0.8 \nl
 E1619 &\phn900 &240 & 22.6 & 27.5 & 21.0 & 26.2 &1.0 & 0.8 \nl
 E1623 &\phn900 &360 & 22.5 & 27.3 & 20.9 & 26.5 &1.0 & 0.8 \nl
\enddata
\end{deluxetable}

\clearpage

\pagestyle{empty}
\begin{deluxetable}{lcccccc}                    
\scriptsize
\tablecaption{Optical Photometric Solutions:
%
%
$m = -2.5*\log{(DN/sec)} +  m_{zp} + X*Airmass + Y*(B-R)$ \label{optphotsoltable}}
\tablehead{
\colhead{UT Date}      &
\colhead{Filter}      & 
\colhead{$N_{stars}$} &
\colhead{$m_{zp}$}      &
\colhead{X}  &
\colhead{Y}  &
\colhead{$\sigma_m$}\nl
}
\startdata
970923  & $B$   & 45 & 24.158   & -0.194        &  0.046        & 0.015 \nl
970924  & $B$   & 29 & 24.158   & -0.204        &  0.046        & 0.016 \nl
980330  & $B$   & 38 & 24.114   & -0.237        &  0.044        & 0.016 \nl
980331  & $B$   & 36 & 24.093   & -0.231        &  0.044        & 0.020 \nl
                &               &       &               &                       &                       & \nl
970923  & $R$   & 38 & 24.501   & -0.069        & -0.003        & 0.012 \nl
970924  & $R$   & 38 & 24.501   & -0.086        & -0.003        & 0.015 \nl
980330  & $R$   & 46 & 24.439   & -0.095        & -0.002        & 0.020 \nl
980331  & $R$   & 57 & 24.438   & -0.093        & -0.002        & 0.020 \nl
\enddata
\end{deluxetable}

\vfill
\clearpage

\pagestyle{empty}
\begin{deluxetable}{rccccc}                                
\scriptsize
\tablecaption{Infrared Observations\label{irobstable}}
\tablehead{
\colhead{ FGC }                 &
\colhead{ Exp Time }            &   
\colhead{ Date }                &
\colhead{ FWHM$_{major}$ ($K_s$) }  &   
\colhead{${a/b}_{\rm PSF}$}     &
\colhead{ PA$_{\rm PSF}$ }      
\nl
\colhead{  }               &   
\colhead{ (min) }               &   
\colhead{ (UT) }               &   
\colhead{ ($^{\prime\prime}$) }               &   
\colhead{  }               &   
\colhead{ (degrees) }                             \nl
}
\startdata
31   & 48       & 970921        & 1.1 & 1.5$\pm$0.6 & -85$\pm$35  \nl
36   & 16:      & 971011        & 0.7 & 1.1$\pm$0.1 & -53$\pm$39 \nl
51   & 22:      & 971011        & 1.3 & 1.7$\pm$0.1 & 46$\pm$37 \nl
84   & 36       & 970922        & 0.7 & 1.1$\pm$0.1 & -52$\pm$14        \nl
130  & 12:      & 971011        &         &                 &                   \nl
     & \phn8& 971012    & (1.1) &       (1.1$\pm$0.1) & (77$\pm$77)     \nl
143  & 20::     & 971011        &         &                 &                   \nl
     & 24       & 971012        & 1.0 & 1.1$\pm$0.1 & -73$\pm$100       \nl
164  & 24       & 971012        & 1.1 & 1.1$\pm$0.1 & 47$\pm$17 \nl
215  & 18       & 970922        &       - &      -          &           -       \nl
225  & 30       & 971012        & (1.2) & (1.2$\pm$0.2) & (18$\pm$18)   \nl
     & 20       & 971015        &         &                 &                   \nl
227  & 24       & 970921        &       - &      -          &           -       \nl
     & 12       & 971015        &         &                 &                   \nl
256  & 24::     & 971011        &         &                 &                   \nl
     & 38       & 971012        & 1.1 & 1.1$\pm$0.1 & 81$\pm$80 \nl
     & 36       & 971015        &         &                 &                   \nl
277  & 30       & 970921        & 1.1 & 1.1$\pm$0.1 & -84$\pm$86\nl
     & 30       & 971015        &         &                 &                   \nl
310  & 24       & 970921        & 0.8 & 1.7$\pm$0.7 & 71$\pm$23 \nl
349  & 38:      & 971011        & 0.9 & 1.1$\pm$0.1 & -70$\pm$20                \nl
     & 12       & 971012        &         &                 &                   \nl
395  & 24       & 970921        & 0.8 & 1.5$\pm$0.4 & 34$\pm$10 \nl
     & 18       & 971015        &         &                 &                   \nl
436  & 24       & 970922        & 1.0 & 1.1$\pm$0.1 & -63$\pm$46        \nl
442  & 12:      & 971011        &         &                 &                   \nl
     & 24       & 971012        & 1.0 & 1.5$\pm$0.7 &  85$\pm$69        \nl
446  & 12       & 970922        & (0.8) & (1.2$\pm$0.2) & (-59$\pm$59)          \nl
780  & 24       & 980415        & 1.2 & 1.2$\pm$0.2 & -33$\pm$15                \nl
901  & 18::     & 980414        &         &                 &                   \nl
     & 18       & 980415        &       - &      -          &           -       \nl
913  & 18       & 980415        & 0.9 & 2.0$\pm$0.6 & 86$\pm$58         \nl
979  & 12       & 980415        & 1.0 & 1.0$\pm$0.1 & -26$\pm$23        \nl
1043 & 12       & 980415        & 1.0 & 1.8$\pm$0.6 & -11$\pm$5 \nl
1063 & 18       & 980415        & 1.1 & 1.5$\pm$0.5 & 89$\pm$5  \nl
1285 & 18       & 980415        & 1.0 & 1.0$\pm$0.1 & -52$\pm$5         \nl
1303 & 12:      & 980414        &         &                 &                   \nl
     & 18       & 980415        & 0.9 & 1.2$\pm$0.3 & 78$\pm$19 \nl
1415 & 10       & 980415        & 1.0 & 1.2$\pm$0.1 & -72$\pm$58                \nl
1440 & 10       & 980415        & 1.1 & 2.0$\pm$0.7 & -58$\pm$37                \nl
1642 & 12:      & 980414        &         &                 &                   \nl
     & 20       & 980415        & 0.8 & 1.9$\pm$0.4 & 66$\pm$53 \nl
1863 & 18       & 980415        & 1.2 & 1.5$\pm$0.5 & -52$\pm$55                \nl
1945 & 12:      & 980414        &         &                 &                   \nl
     & 18       & 980415        & 0.9 & 1.5$\pm$0.6 & 83$\pm$61 \nl
1948 & 12:      & 980414        &         &                 &                   \nl
     & 18       & 980415        & 1.1 & 1.9$\pm$0.7 & 82$\pm$72 \nl
1971 & \phn8::  & 980414        &     &             &           \nl
     & 30       & 980415        & 1.1 & 1.6$\pm$0.6 & 59$\pm$47 \nl
2131 & 18       & 980415        & 1.2 & 1.4$\pm$0.1 & 69$\pm$4  \nl
2135 & 18       & 980415        & 1.0 & 1.3$\pm$0.2 & 59$\pm$56 \nl
2217 & 12       & 980415        & 1.0 & 1.4$\pm$0.2 & 12$\pm$4  \nl
2264 & 12       & 970921        & 1.0 & 1.3$\pm$0.2 & 46$\pm$34 \nl
2292 & 24       & 970921        & 0.9 & 1.3$\pm$0.2 & 91$\pm$26 \nl
2367 & 18       & 971011        & 1.1 & 1.1$\pm$0.2 & 87$\pm$49 \nl
     & 38       & 971012        &     &             &           \nl
2369 & 36       & 970921        & 1.0 & 1.2$\pm$0.2 & 59$\pm$75 \nl
2548 & 36       & 970921        &     &             &           \nl
     & 48       & 971012        & 1.2 & 1.2$\pm$0.1 & 28$\pm$7  \nl
2558 & 18:      & 971011        &         &                 &                   \nl
     & 36       & 971012        & 0.5 & 1.1$\pm$0.2 & -26$\pm$20                \nl
E1371 & 18      & 970922        & 1.2 & 1.4$\pm$0.1 & 60$\pm$3  \nl
E1404 & 18      & 971011        &         &                 &                   \nl
      & 30      & 971012        & 0.9 & 1.1$\pm$0.1 & 59$\pm$14 \nl
E1440 & 30      & 970922        & 1.1 & 1.2$\pm$0.2 & -67$\pm$20                \nl
      & 16      & 971015        &         &                 &                   \nl
E1447 & 30      & 970921        & 0.9 & 1.2$\pm$0.1 & -44$\pm$12                \nl
E1498 & 18      & 970922        & 0.8 & 1.1$\pm$0.1 & 74.4$\pm$4.2              \nl
E1619 & 18      & 971011        & 1.1 & 1.2$\pm$0.2     & -66$\pm$46            \nl
E1623 & 18      & 971011        & 0.9 & 1.1$\pm$0.1 & -25$\pm$25                \nl
\enddata
\tablecomments{Single colons (``:'') indicate data taken during clear portions
of non-photometric nights.  Double colons (``::'') indicate taken during
non-photometric conditions.  See discussion in \S\ref{sec:irphot}}.
\end{deluxetable}

\clearpage

\pagestyle{empty}
\begin{deluxetable}{lcccccc}                    
\scriptsize
\tablecaption{Infrared Photometric Solutions:
%
%
$m = -2.5*\log{(DN/sec)} +  m_{zp} + X*Airmass + Y*(J-K_s)$ 
\label{irphotsoltable}}
\tablehead{
\colhead{UT Date}      &
\colhead{Filter}      & \colhead{$N_{stars}$} &
\colhead{$m_{zp}$}      &
\colhead{X}  &
\colhead{Y}  &
\colhead{$\sigma_m$}\nl
}
\startdata
970922  & $K_s$ & 46 & 21.790   & -0.077& -0.067 & 0.023 \nl
970923  & $K_s$ & 25 & 21.790   & -0.072& -0.067 & 0.023 \nl
971011  & $K_s$ & 48 & (21.793)& (-0.065)& (-0.060)& (0.016)\nl
971012  & $K_s$ & 28 & 21.790   & -0.071& -0.067 & 0.023 \nl
971015  & $K_s$ & 28 & 21.790   & -0.076& -0.067 & 0.023 \nl
        &               &    &                  &               &               & \nl
980414  & $K_s$ & 19  & (21.006)& (-0.086)&(-0.067)& (0.027) \nl
980415  & $K_s$ & 46 & 20.950 & -0.058 & -0.067 & 0.045 \nl
\enddata
\end{deluxetable}

\clearpage

\pagestyle{empty}
\begin{deluxetable}{rccccccc}                    
\scriptsize
\tablecaption{Galaxy Magnitudes\label{magtable}}
\tablehead{
\colhead{FGC}      &
\colhead{$m_{R25}(B)$} & \colhead{$m_{R25}(R)$} & \colhead{$m_{R25}(K)$} &
\colhead{$m_{B27}(B)$} & \colhead{$m_{R26}(R)$} & \colhead{$m_{K22}(K)$} &
\colhead{$E(B-V)$}\nl
}
\startdata
   31 & 17.58$\pm$0.02 & 16.90$\pm$0.03 & 15.15$\pm$0.46 & 17.52$\pm$0.03 & 16.83$\pm$0.03 & 15.42$\pm$0.27 & 0.053\nl
   36 & 17.45$\pm$0.02 & 16.32$\pm$0.01 & 13.84$\pm$0.18 & 17.43$\pm$0.02 & 16.30$\pm$0.01 & 13.76$\pm$0.21 & 0.123\nl
   51 & 16.64$\pm$0.02 & 15.84$\pm$0.02 & 13.79$\pm$0.51 & 16.54$\pm$0.03 & 15.72$\pm$0.03 & 13.90$\pm$0.32 & 0.107\nl
   84 & 17.57$\pm$0.02 & 16.69$\pm$0.02 & 14.60$\pm$0.17 & 17.52$\pm$0.03 & 16.64$\pm$0.02 & 14.76$\pm$0.11 & 0.029\nl
  130 & 17.73$\pm$0.04 & 16.20$\pm$0.03 & 12.92$\pm$0.05 & 17.72$\pm$0.04 & 16.18$\pm$0.03 & 12.86$\pm$0.07 & 0.049\nl
  143 & 17.50$\pm$0.07 & 16.45$\pm$0.08 & 14.42$\pm$0.24 & 17.43$\pm$0.07 & 16.37$\pm$0.07 & 14.70$\pm$0.17 & 0.049\nl
  164 & 17.89$\pm$0.02 & 17.19$\pm$0.02 & 15.24$\pm$0.28 & 17.80$\pm$0.03 & 17.09$\pm$0.03 & 15.56$\pm$0.17 & 0.029\nl
  215 & 17.23$\pm$0.02 & 16.11$\pm$0.02 & 13.40$\pm$0.15 & 17.20$\pm$0.02 & 16.09$\pm$0.02 & 13.33$\pm$0.16 & 0.051\nl
  225 & 17.08$\pm$0.02 & 16.21$\pm$0.02 & 14.11$\pm$0.19 & 17.05$\pm$0.02 & 16.17$\pm$0.03 & 14.18$\pm$0.14 & 0.049\nl
  227 & 17.22$\pm$0.03 & 15.93$\pm$0.02 & 13.37$\pm$0.08 & 17.19$\pm$0.04 & 15.88$\pm$0.03 & 13.39$\pm$0.07 & 0.112\nl
  256 & 18.08$\pm$0.02 & 17.24$\pm$0.01 & 15.24$\pm$0.21 & 18.01$\pm$0.02 & 17.18$\pm$0.02 & 15.47$\pm$0.16 & 0.050\nl
  277 & 18.01$\pm$0.03 & 16.60$\pm$0.02 & 14.10$\pm$0.14 & 17.94$\pm$0.03 & 16.52$\pm$0.03 & 14.10$\pm$0.12 & 0.201\nl
  310 & 17.68$\pm$0.04 & 16.20$\pm$0.04 & 13.40$\pm$0.10 & 17.65$\pm$0.05 & 16.17$\pm$0.04 & 13.35$\pm$0.11 & 0.192\nl
  349 & 17.46$\pm$0.02 & 16.37$\pm$0.02 & 13.99$\pm$0.07 & 17.44$\pm$0.02 & 16.35$\pm$0.02 & 14.00$\pm$0.07 & 0.152\nl
  395 & 17.74$\pm$0.03 & 16.32$\pm$0.03 & 13.48$\pm$0.17 & 17.71$\pm$0.04 & 16.28$\pm$0.03 & 13.50$\pm$0.11 & 0.126\nl
  436 & 17.39$\pm$0.04 & 16.05$\pm$0.03 & 13.30$\pm$0.09 &      $-$       & 15.99$\pm$0.04 & 13.31$\pm$0.08 & 0.330\nl
  442 & 17.26$\pm$0.02 & 16.12$\pm$0.02 & 13.74$\pm$0.10 & 17.22$\pm$0.02 & 16.10$\pm$0.02 & 13.73$\pm$0.09 & 0.144\nl
  446 & 16.17$\pm$0.02 & 14.68$\pm$0.02 & 11.39$\pm$0.13 & 16.15$\pm$0.02 & 14.67$\pm$0.02 & 11.36$\pm$0.14 & 0.234\nl
  780 & 15.77$\pm$0.07 & 14.88$\pm$0.07 & 12.85$\pm$0.55 & 15.70$\pm$0.09 & 14.83$\pm$0.07 & 12.85$\pm$0.39 & 0.035\nl
  901 & 17.48$\pm$0.03 & 16.37$\pm$0.03 & 13.92$\pm$0.23 & 17.44$\pm$0.04 & 16.34$\pm$0.03 & 13.92$\pm$0.22 & 0.022\nl
  913 & 17.04$\pm$0.02 & 16.10$\pm$0.02 & 13.95$\pm$0.26 & 17.02$\pm$0.02 & 16.08$\pm$0.02 & 13.91$\pm$0.22 & 0.038\nl
  979 & 15.42$\pm$0.02 & 14.36$\pm$0.02 & 11.87$\pm$0.12 & 15.39$\pm$0.02 & 14.35$\pm$0.02 & 11.81$\pm$0.14 & 0.057\nl
 1043 & 15.37$\pm$0.02 & 14.22$\pm$0.02 & 11.41$\pm$0.15 & 15.33$\pm$0.03 & 14.20$\pm$0.03 & 11.41$\pm$0.12 & 0.010\nl
 1063 & 17.07$\pm$0.02 & 16.29$\pm$0.02 & 14.15$\pm$0.26 & 17.03$\pm$0.03 & 16.25$\pm$0.03 & 14.20$\pm$0.21 & 0.032\nl
 1285 & 14.75$\pm$0.02 & 13.75$\pm$0.02 & 11.87$\pm$2.19 & 14.73$\pm$0.02 & 13.72$\pm$0.03 &      $-$      & 0.013\nl
 1303 & 17.38$\pm$0.02 & 16.51$\pm$0.02 & 14.35$\pm$0.62 & 17.30$\pm$0.03 & 16.45$\pm$0.02 & 13.75$\pm$0.69 & 0.013\nl
 1415 & 15.19$\pm$0.02 & 14.21$\pm$0.02 & 12.05$\pm$0.54 & 15.16$\pm$0.02 & 14.18$\pm$0.03 & 12.04$\pm$0.37 & 0.023\nl
 1440 & 15.75$\pm$0.02 & 14.48$\pm$0.02 & 11.52$\pm$0.15 & 15.72$\pm$0.03 & 14.46$\pm$0.02 &      $-$      & 0.020\nl
 1642 & 16.79$\pm$0.04 & 15.79$\pm$0.06 & 13.91$\pm$0.57 & 16.75$\pm$0.05 &      $-$      & 13.97$\pm$0.37 & 0.033\nl
 1863 & 15.28$\pm$0.14 & 14.44$\pm$0.16 & 12.53$\pm$0.93 & 15.24$\pm$0.15 & 14.40$\pm$0.16 &      $-$      & 0.051\nl
 1945 & 16.66$\pm$0.03 & 15.68$\pm$0.03 & 13.34$\pm$0.46 & 16.59$\pm$0.03 & 15.63$\pm$0.03 & 13.07$\pm$0.44 & 0.075\nl
 1948 & 16.99$\pm$0.02 & 16.12$\pm$0.02 & 14.10$\pm$0.77 & 16.88$\pm$0.08 & 16.07$\pm$0.03 & 14.11$\pm$0.56 & 0.088\nl
 1971 & 16.70$\pm$0.02 & 15.30$\pm$0.02 & 12.54$\pm$0.28 & 16.64$\pm$0.02 & 15.25$\pm$0.02 &      $-$      & 0.042\nl
 2131 & 16.73$\pm$0.05 & 15.52$\pm$0.06 & 13.01$\pm$0.38 & 16.70$\pm$0.06 & 15.48$\pm$0.06 & 12.91$\pm$0.40 & 0.167\nl
 2135 & 17.60$\pm$0.03 & 16.40$\pm$0.03 & 13.82$\pm$0.32 & 17.57$\pm$0.04 & 16.36$\pm$0.04 &      $-$      & 0.118\nl
 2217 & 15.77$\pm$0.27 & 13.75$\pm$0.26 &  9.86$\pm$1.40 & 15.74$\pm$0.28 & 13.72$\pm$0.26 &      $-$      & 0.294\nl
 2264 & 16.89$\pm$0.24 & 15.50$\pm$0.20 & 12.26$\pm$0.35 &      $-$       &      $-$      & 12.27$\pm$0.21 & 0.225\nl
 2292 & 17.73$\pm$0.15 & 16.78$\pm$0.16 & 14.16$\pm$1.13 & 17.58$\pm$0.17 & 16.71$\pm$0.16 &      $-$      & 0.100\nl
 2367 & 17.90$\pm$0.02 & 17.02$\pm$0.03 & 14.90$\pm$0.22 & 17.84$\pm$0.02 & 16.96$\pm$0.03 & 15.02$\pm$0.16 & 0.067\nl
 2369 & 17.71$\pm$0.03 & 16.76$\pm$0.02 & 14.70$\pm$0.20 & 17.67$\pm$0.04 & 16.72$\pm$0.03 & 14.81$\pm$0.15 & 0.068\nl
 2548 & 17.34$\pm$0.08 & 16.26$\pm$0.06 & 13.97$\pm$0.15 & 17.29$\pm$0.10 & 16.21$\pm$0.07 & 14.06$\pm$0.12 & 0.089\nl
 2558 & 16.70$\pm$0.02 & 15.73$\pm$0.02 & 13.58$\pm$0.16 & 16.68$\pm$0.02 & 15.70$\pm$0.02 & 13.61$\pm$0.12 & 0.039\nl
E1371 & 17.96$\pm$0.15 & 16.17$\pm$0.15 & 12.59$\pm$0.14 & 17.94$\pm$0.16 & 16.13$\pm$0.16 & 12.57$\pm$0.15 & 0.247\nl
E1404 & 17.82$\pm$0.06 & 16.63$\pm$0.06 & 14.08$\pm$0.14 & 17.79$\pm$0.07 & 16.59$\pm$0.07 & 14.07$\pm$0.14 & 0.169\nl
E1440 & 17.84$\pm$0.06 & 16.70$\pm$0.06 & 14.30$\pm$0.32 & 17.78$\pm$0.06 & 16.65$\pm$0.07 & 14.35$\pm$0.22 & 0.146\nl
E1447 & 18.62$\pm$0.02 & 16.93$\pm$0.02 & 13.63$\pm$0.08 & 18.59$\pm$0.03 & 16.89$\pm$0.02 & 13.59$\pm$0.10 & 0.057\nl
E1498 & 17.71$\pm$0.02 & 16.31$\pm$0.02 & 13.36$\pm$0.08 & 17.69$\pm$0.02 & 16.28$\pm$0.02 & 13.35$\pm$0.08 & 0.035\nl
E1619 & 17.75$\pm$0.02 & 16.05$\pm$0.02 & 12.70$\pm$0.08 & 17.72$\pm$0.02 & 16.00$\pm$0.03 & 12.66$\pm$0.09 & 0.043\nl
E1623 & 18.24$\pm$0.04 & 16.88$\pm$0.03 & 13.73$\pm$0.08 & 18.17$\pm$0.05 & 16.83$\pm$0.03 & 13.67$\pm$0.10 & 0.045\nl
\enddata
\end{deluxetable}

\clearpage

\section{Figure Captions}

\figcaption[Dalcanton.f1.ps]{ The distribution of morphological 
T-types (upper) and surface brightness classes (lower) for our
subsample of the FGC (shaded histogram) and the entire FGC (dotted
line, scaled vertically to fit the display).  The morphological
classes and surface brightness classes are those listed in the FGC.
The surface brightness classes are defined such that "I" is the
highest surface brightness and "IV" is the lowest.  We have made no
correction for the slightly different surface brightness limits of the
FGC and FGCE.
\label{sbtypefig}}

\figcaption[Dalcanton.f2.ps]{ The distribution of blue axial ratios 
for our subsample of the FGC (shaded histrogram) and the entire FGC
(dotted line, scaled vertically to fit the display).  We have made no
correction for the slightly different depths of the FGC and FGCE.
\label{axialratiofig}}

\figcaption[Dalcanton.f3.ps]{ $B$ (lower left), $R$ (upper), and $K_s$
  (lower right) images and contour diagrams of the 49 galaxies of our
  sample.  All images are displayed with the same stretch and display
  range, such that apparent surface brightness variations from
  galaxy-to-galaxy reflect true variations in the surface brightness.
  The contour levels are separated by $1\surfb$, with the dark
  reference contour drawn at $\mu_B=27.0\surfb$, $\mu_R=26\surfb$, and
  $\mu_{K_s}=21\surfb$ in the lower left, upper, and lower right
  contour images respectively.  The solid circle in the upper right of
  each image has a diameter equal to the FWHM of the point-spread
  function for that image.  The horizontal line in the lower left of
  the upper $R$-band image is equal to $1\,h^{-1}_{50}\kpc$ if the
  galaxy is at a distance of $V_\odot/H_0$; we have taken the
  recessional velocity $V_\odot$ from the published single dish HI
  measurements described in \S\ref{samplesec} (if available) or from
  our long slit H$\alpha$ spectroscopy.  At isophotal levels where
  there is sufficient signal-to-noise, contours are defined from the
  unsmoothed image.  However, fainter isophote are defined using a
  smoothed version of the image (see \S\ref{magsec}).  Tick marks are
  given at every 5 arcseconds.  The light grey shading under the
  contours indicates regions which were masked. {\bf THESE IMAGES ARE
    AVAILABLE IN POSTSCRIPT FORMAT AT} {\tt
    ftp://ftp.astro.washington.edu/pub/users/jd/FGC/dalcanton.f3.ps.gz}
  {\bf or in low-resolution JPEG format from astro-ph}
\label{contourfig}}

\figcaption[Dalcanton.f4a.ps,Dalcanton.f4b.ps,Dalcanton.f4c.ps,
Dalcanton.f4d.ps,Dalcanton.f4e.ps]{ The standard deviation of unmasked
sky pixels in the final $B$ (heavy line), $R$ (medium line) and $K_s$
(light line, plotted as $\mu+5$) images, as a function of the boxcar
smoothing length.  The dashed line gives the expected relation for a
perfectly flat, uniform, Poisson sky background.  Only points where
there are at least 25 independent, unmasked regions in the smoothed
images are plotted.
\label{noisefig}}

\vfill
\clearpage

\begin{figure}[p]
\centerline{ \psfig{figure=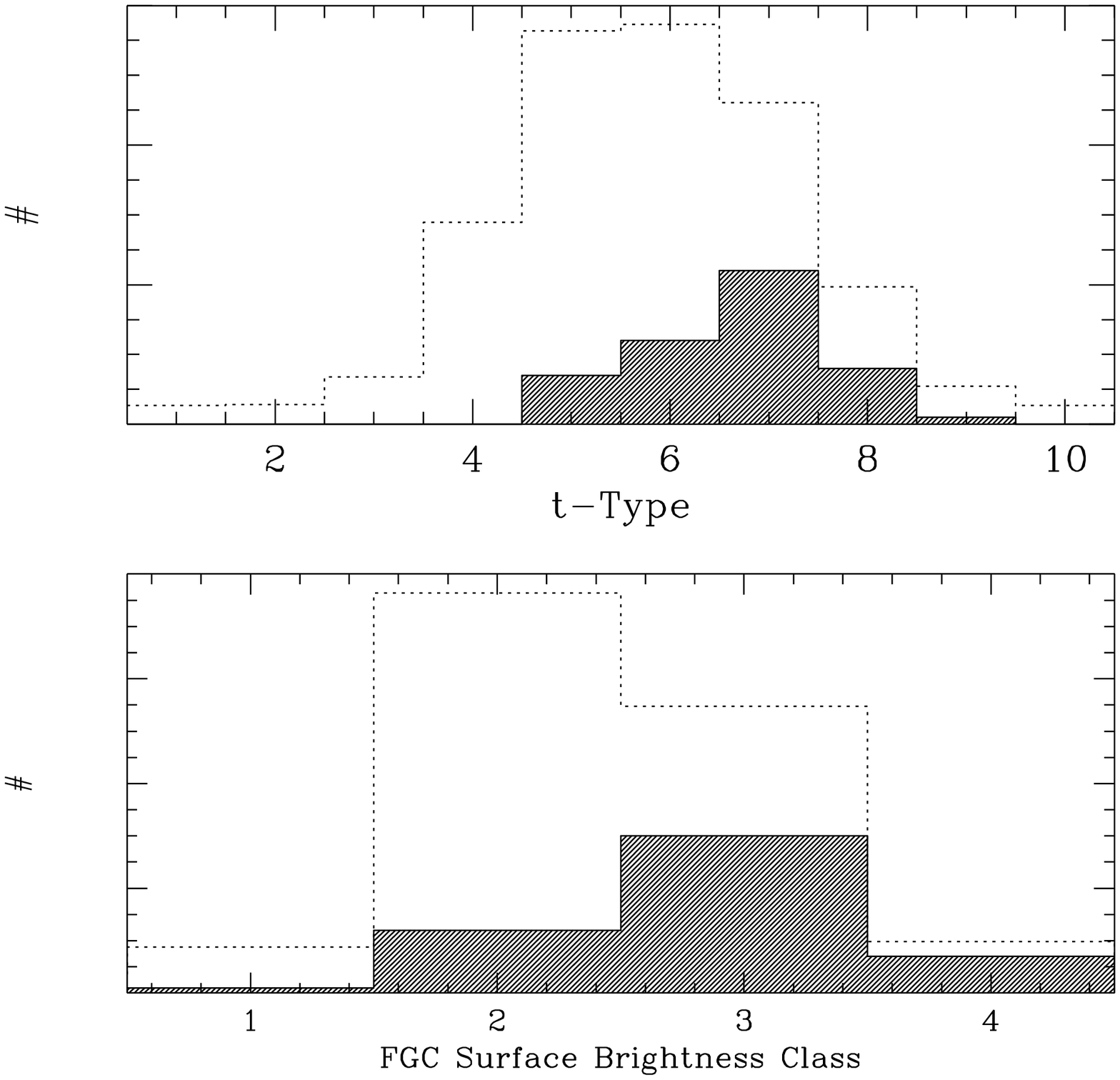} }
\begin{flushright}{\bigskip\cap Figure \ref{sbtypefig}}\end{flushright}
\end{figure}
\vfill
\clearpage

\begin{figure}[p]
\centerline{ \psfig{figure=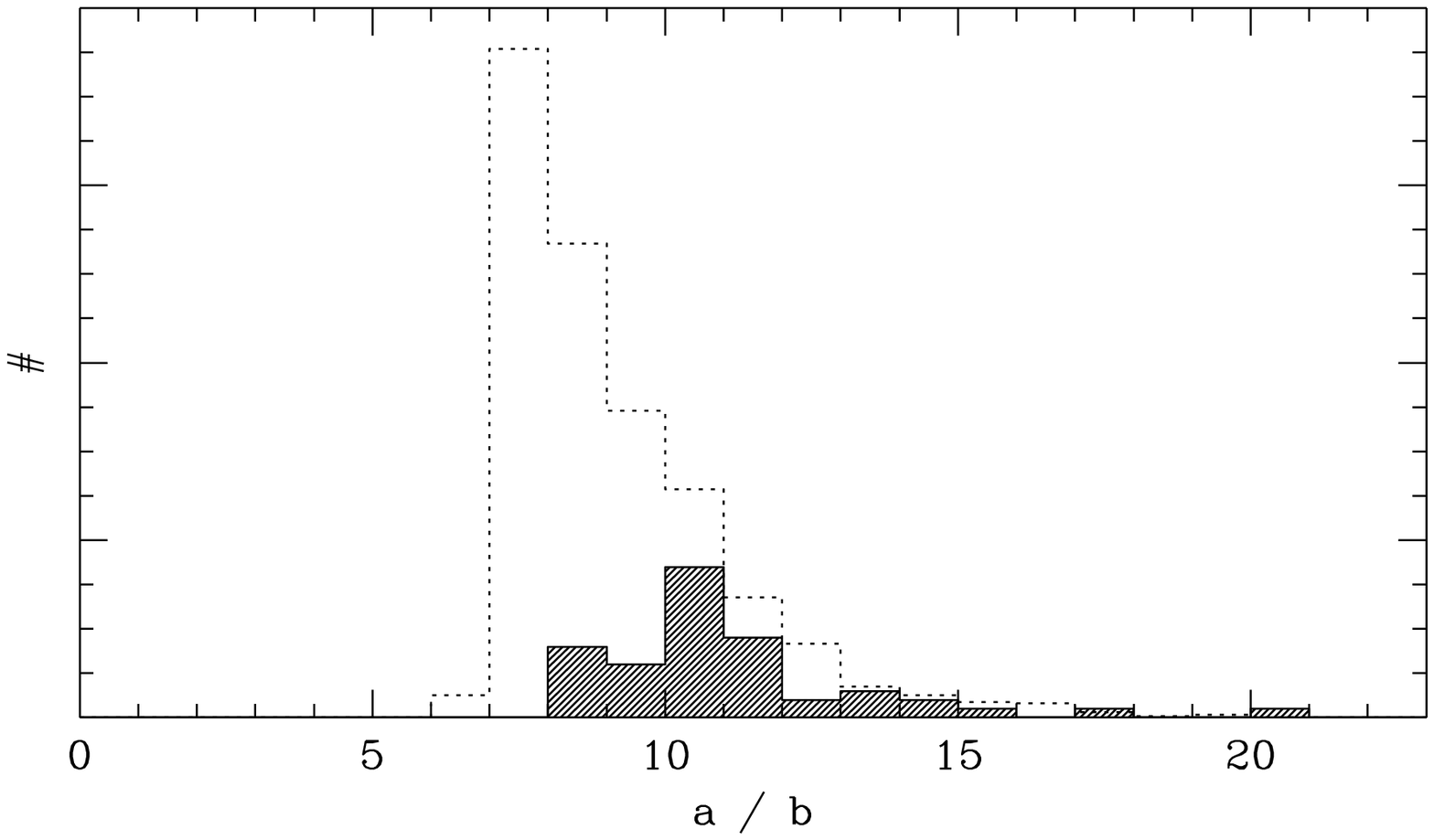} }
\begin{flushright}{\bigskip\cap Figure \ref{axialratiofig}}\end{flushright}
\end{figure}
\vfill
\clearpage


\begin{figure}[p]
\centerline{ \psfig{figure=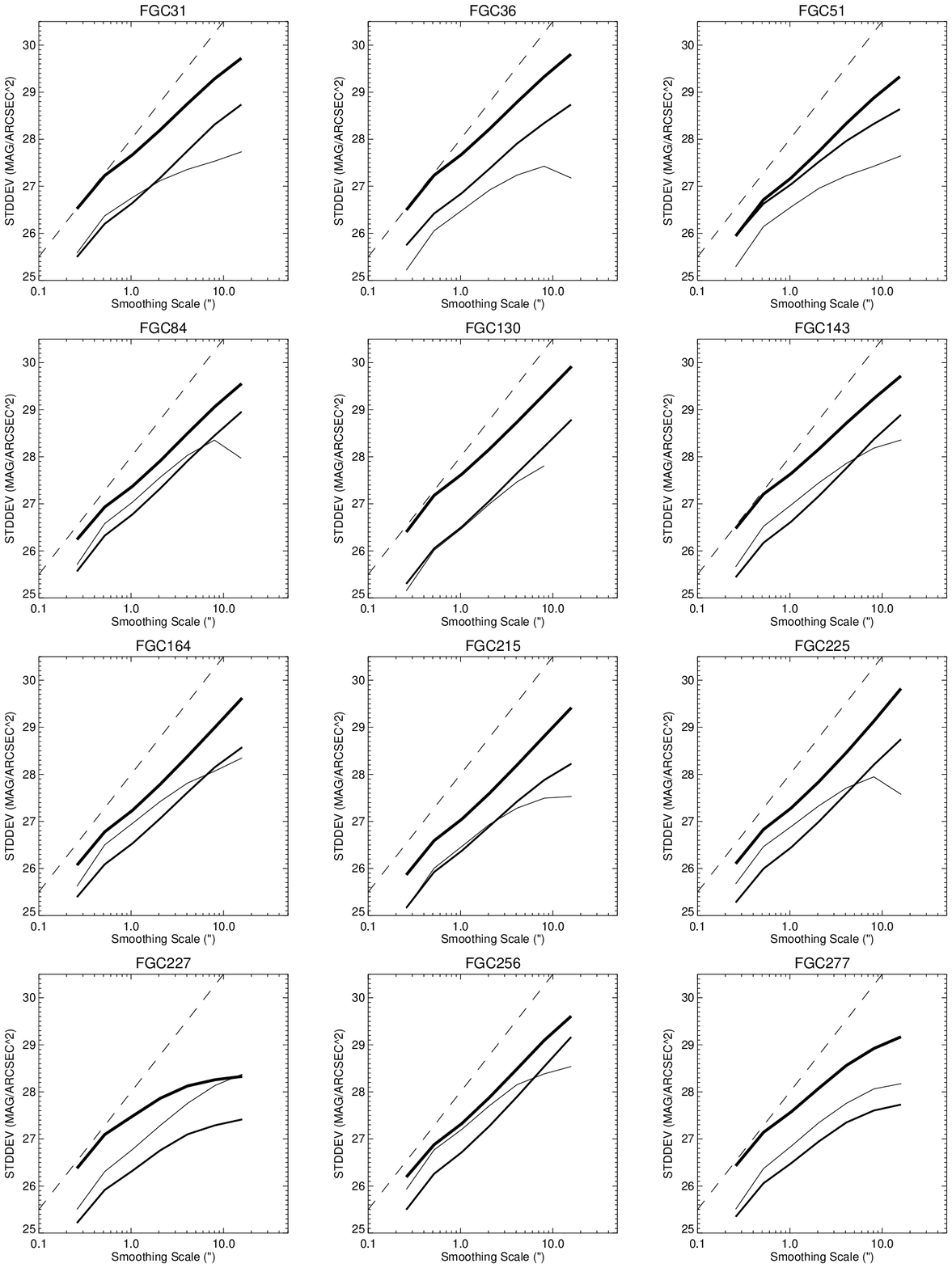,height=8.5in} }
\begin{flushright}{\bigskip\cap Figure \ref{noisefig}}\end{flushright}
\end{figure}
\vfill
\clearpage

\begin{figure}[p]
\centerline{ \psfig{figure=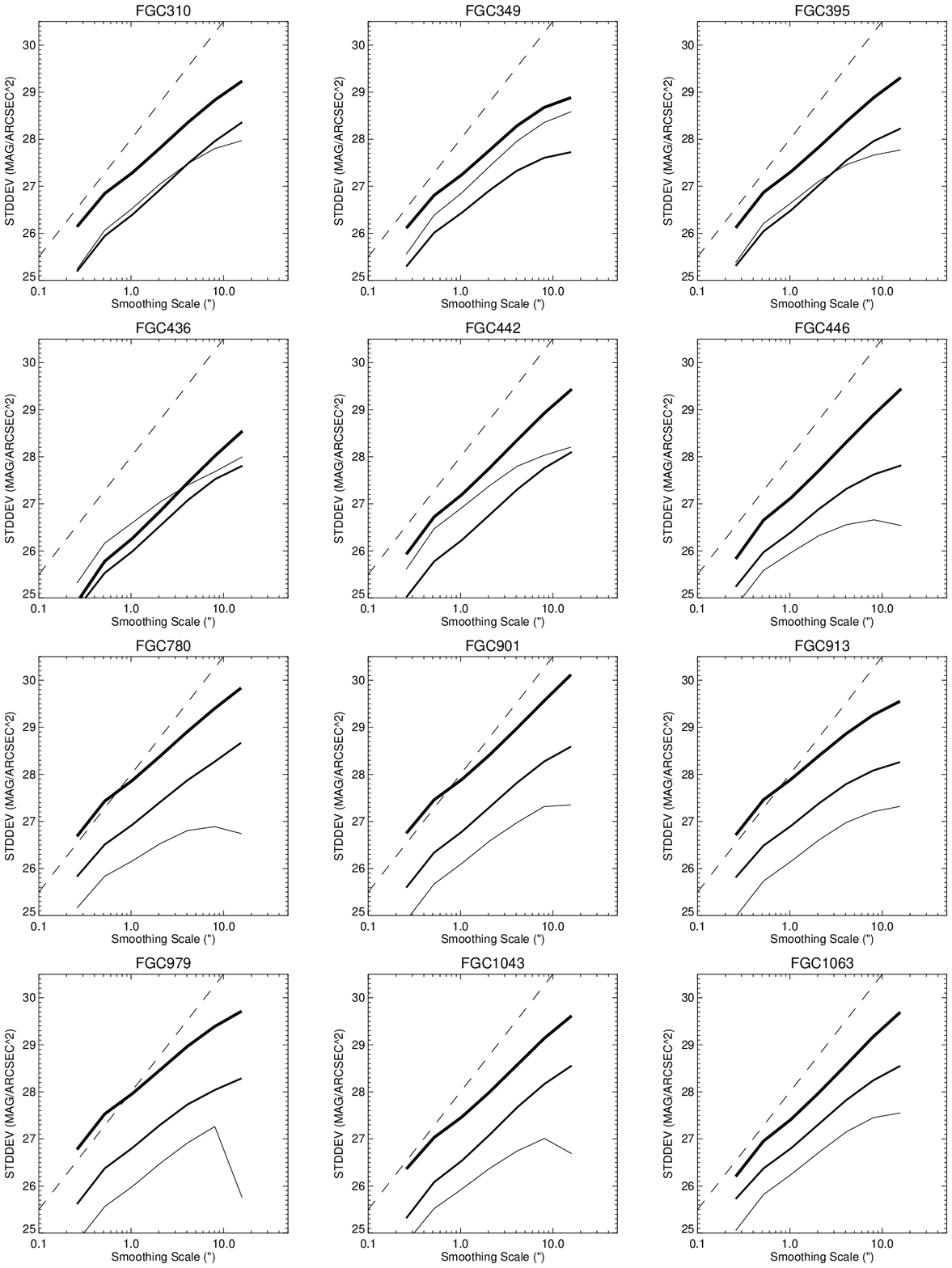,height=8.5in} }
\begin{flushright}{\bigskip\cap Figure \ref{noisefig} (continued)}\end{flushright}
\end{figure}
\vfill
\clearpage

\begin{figure}[p]
\centerline{ \psfig{figure=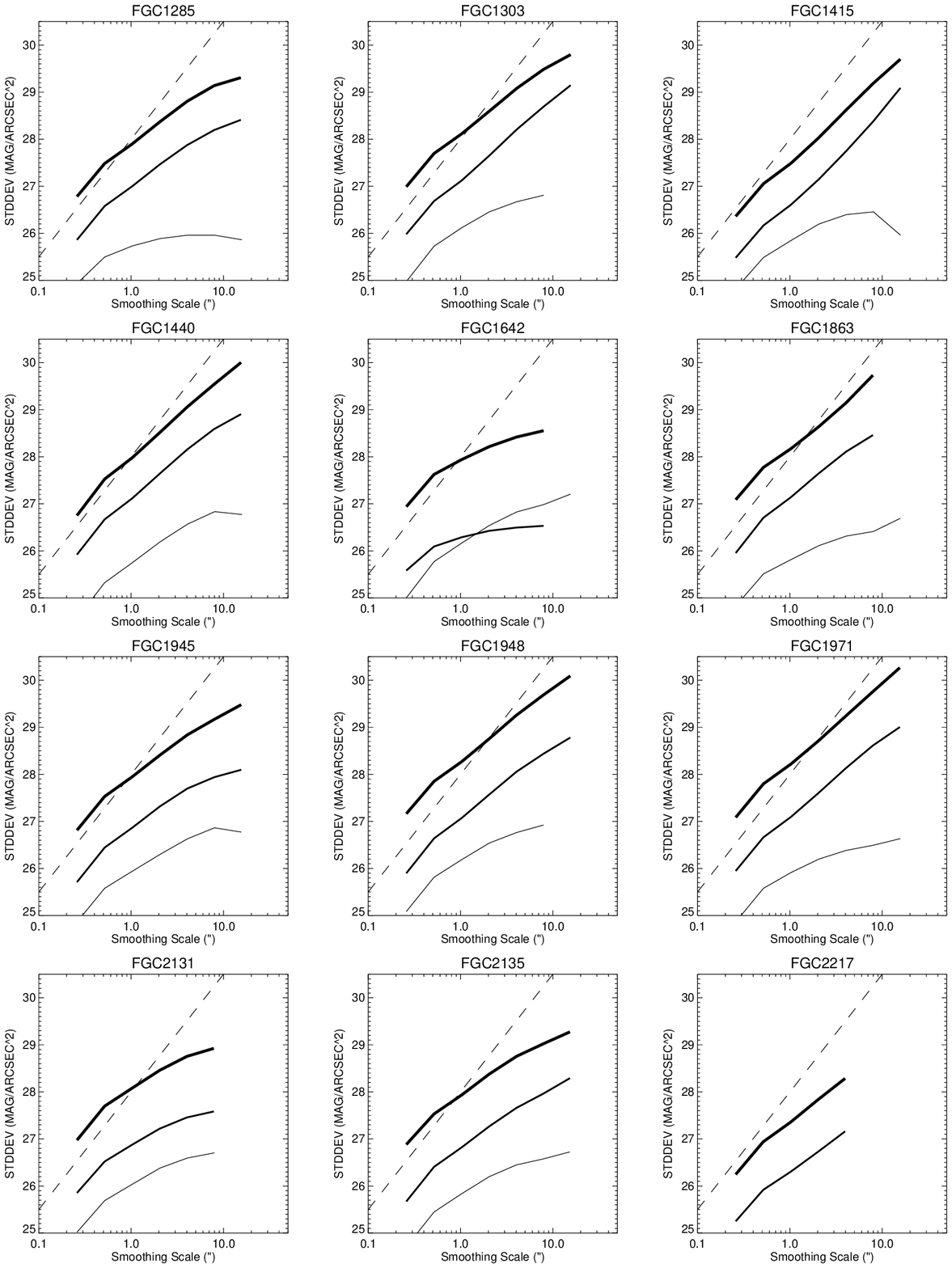,height=8.5in} }
\begin{flushright}{\bigskip\cap Figure \ref{noisefig} (continued)}\end{flushright}
\end{figure}
\vfill
\clearpage

\begin{figure}[p]
\centerline{ \psfig{figure=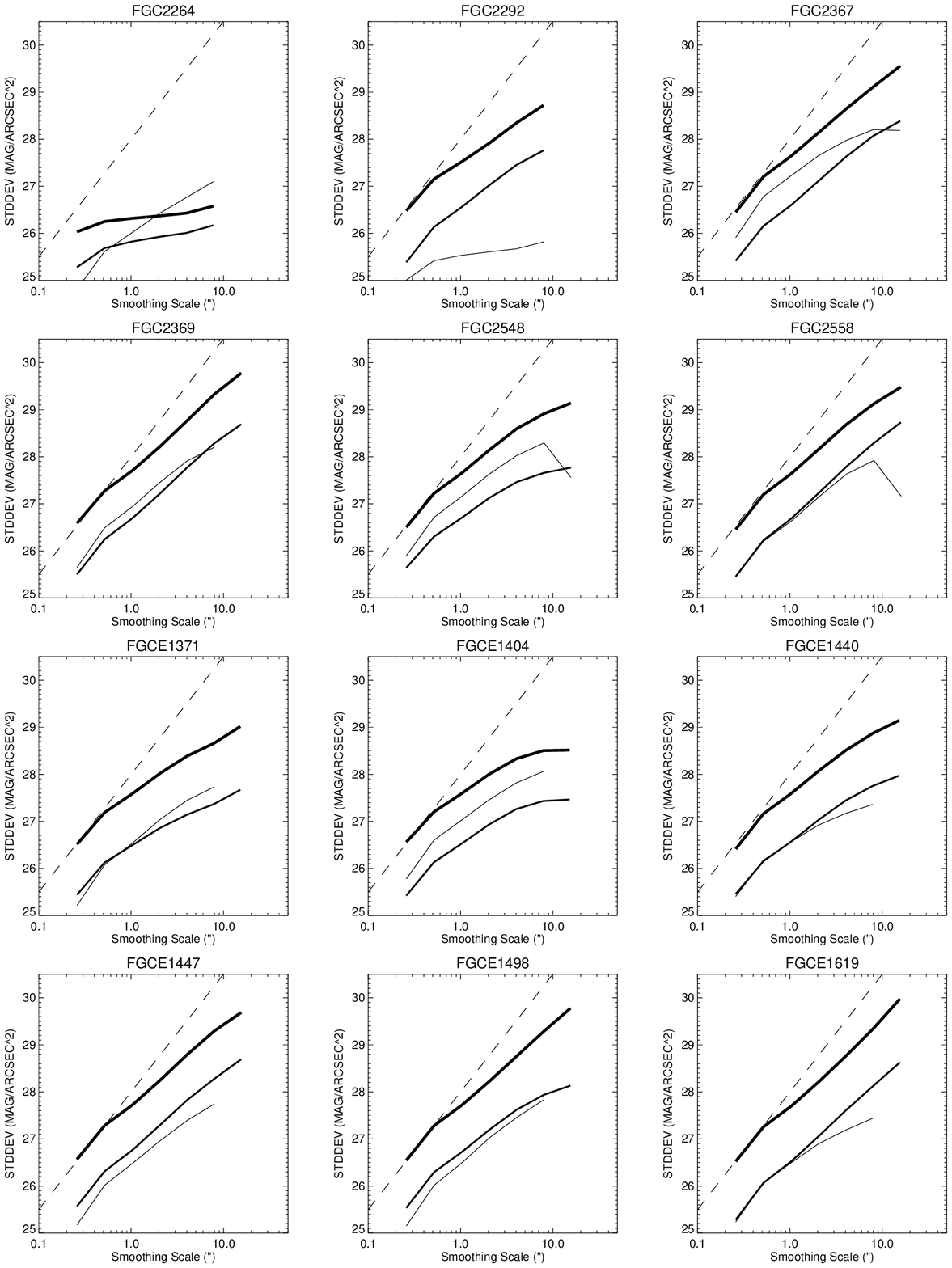,height=8.5in} }
\begin{flushright}{\bigskip\cap Figure \ref{noisefig} (continued)}\end{flushright}
\end{figure}
\vfill
\clearpage

\begin{figure}[p]
\centerline{ \psfig{figure=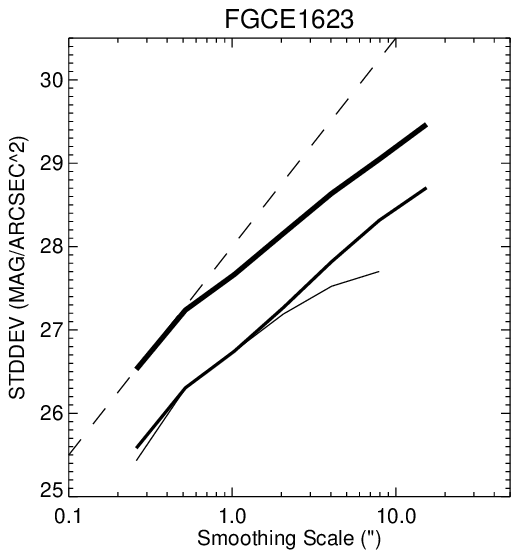,height=8.5in} }
\begin{flushright}{\bigskip\cap Figure \ref{noisefig} (continued)}\end{flushright}
\end{figure}
\vfill
\clearpage

\end{document}